\documentclass[fleqn,usenatbib]{mnras}
\usepackage{newtxtext,newtxmath}
\usepackage[T1]{fontenc}

\DeclareRobustCommand{\VAN}[3]{#2}
\let\VANthebibliography\thebibliography
\def\thebibliography{\DeclareRobustCommand{\VAN}[3]{##3}\VANthebibliography}
\setcitestyle{citesep={,}}

\usepackage{graphicx,float, caption}

\usepackage{wrapfig}
\restylefloat{table}
\usepackage[tableposition = top]{caption}
\bibpunct{(}{)}{;}{a}{}{,}
\usepackage{newtxtext,newtxmath}
\newcommand{\mpyr}{$\text{M}_{\odot}\text{yr}^{-1}$}
\newcommand{\mpyrkpc}{$\text{M}_{\odot}\text{yr}^{-1}\text{kpc}^{-2}$}

\title[Deriving the intrinsic properties of M51]{Deriving the intrinsic properties of M51 with radiative transfer models}

\author[C. J. Inman et al.]{
Christopher J. Inman,$^{1}$\thanks{E-mail: CJInman@uclan.ac.uk}
Cristina C. Popescu,$^{1,2}$\thanks{E-mail: CPopescu@uclan.ac.uk}
Mark Rushton $^{3}$
and David Murphy$^{1}$\\
$^1$University of Central Lancashire, Jeremiah Horrocks Institute, Preston, PR1 2HE, UK \\
$^2$ Max Planck Institut f\"ur Kernphysik, Saupfercheckweg 1, 69117 Heidelberg, Germany\\
$^{3}$The Astronomical Institute of the Romanian Academy, Str. Cutitul de Argint 5, Bucharest, Romania
}

\date{Accepted XXX. Received YYY; in original form ZZZ}

\pubyear{2023}
\begin{document}
\label{firstpage}
\pagerange{\pageref{firstpage}--\pageref{lastpage}}
\maketitle

\begin{abstract}
A quantitative derivation of the intrinsic properties of galaxies related to their fundamental building blocks, gas, dust and stars is essential for our understanding of galaxy evolution. A fully self-consistent derivation of these properties can be achieved with radiative transfer (RT) methods that are constrained by panchromatic imaging observations. Here we present an axi-symmetric RT model of the UV-optical-FIR/submm spectral and spatial energy distribution of the face-on spiral galaxy M51. The model reproduces 
reasonably well the azimuthally averaged radial profiles derived from the imaging data available for this galaxy, from GALEX, SDSS, 2MASS, Spitzer and Herschel. We model the galaxy with three distinct morphological components: a bulge, an inner disc and a main disc. We derive the length parameters of the stellar emissivity and of the dust distribution. We also derive the intrinsic global and spatially resolved parameters of M51. We find a faint \lq\lq outer disc\rq\rq\ bridging M51 with its companion galaxy M51b. Finally, we present and discuss an alternative model, with dust properties that change within the galaxy.
\end{abstract}

\begin{keywords}
radiative transfer -- galaxies: disc -- galaxies: stellar content -- galaxies: structure -- ISM: dust, extinction -- galaxies: spiral
\end{keywords}

%%%%%%%%%%%%%%%%%%%%%%%%%%%%%%%%%%%%%%%%%%%%%%%%%%
\defcitealias{Popescu2011}{PT11}
%%%%%%%%%%%%%%%%% BODY OF PAPER %%%%%%%%%%%%%%%%%%

\section{Introduction}
\label{section:intro}
Dust is prevalent within the interstellar medium (ISM) of star-forming galaxies \citep{Trumpler1930}, and, although it constitutes only $1\%$ of the mass of this ISM \citep{Greenberg1963}, it plays an important role in the thermodynamic balance of galaxies \citep{Montier2004, Giard2008, Popescu2010,Natale2010}, since it regulates the cooling and heating mechanisms of the interstellar gas \citep{Dwek&Werner1981}. 

The interstellar dust also affects our view of galaxies.
This is because stellar photons, primarily in the ultraviolet (UV)/optical range, are continuously absorbed and scattered by the interstellar dust. 
Photons that are scattered can travel long distances within the ISM before they are scattered again, absorbed or leave the galaxy.
Due to scattering, what would be otherwise an isotropic process related to the emission of stellar light in galaxies, becomes highly anisotropic.
As such, along any line of sight through a galaxy, the stellar light may be diminished due to extinction, but also amplified by stellar photons being scattered into the line of sight. Thus, for a extended distribution of emitters and absorbers, the complex effect of absorption and scattering processes is what we call attenuation.

The attenuation is dependent not only on the optical properties of the dust grains (as is the case for the extinction processes), but also on the relative distribution of stars and dust, and on the orientation of a galaxy with respect to the observer \citep{Byun1994,Bianchi1996,Kuchinski1998,Ferrara1999,Baes2001,Tuffs2004,Pierini2004,Fischera2004,Natale2015}. Dust attenuation affects the observed spatially integrated stellar luminosity of a galaxy, but also its surface brightness distribution and the geometric parameters derived from surface-brightness photometry, like scalelength  of discs, effective radii of bulges and Sérsic indexes: \cite{Byun1994,Evans1994,Cunow2001,Mollenhoff2006,Gadotti2010,Pastrav2013a,Pastrav2013b,Savchenko2023}.

Due to dust attenuation the observed UV/optical images of star forming galaxies are far from providing a direct tracer of the constituent stellar populations distributed throughout the galaxy. In order to derive the intrinsic distribution of stars of all ages, and thus to understand how stars form and evolve in galaxies, one also needs to jointly derive the distribution of dust in galaxies \citep{Popescu2010}. This can be done by considering both the absorption and emission properties of interstellar dust. Thus, the stellar light that is absorbed by dust is re-emitted in the infrared (IR) and the energy balance between direct and re-radiated stellar light is an important constraint on the modelling of transfer of radiation in galaxies. In particular dust emission contains unique information on both the spatial and spectral energy distribution (SED) of the energy density of the radiation fields that heat the dust grains, and on the dust properties and dust opacity. Accounting for both dust attenuation and dust emission is thus needed to provide a self-consistent model of a galaxy.

Galaxy models that are based on radiative transfer (RT) methods (\citealp{Kylafis2005}; \citealp{Steinacker2013}) and constrained by panchromatic imaging observations are the most powerful tools in quantifying the attenuation of stellar light in galaxies and in deriving the intrinsic parameters of galaxies \citep{Popescu2021}. This type of modelling was initially performed on edge-on galaxies, since in this orientation it is possible to directly constrain the vertical distribution of stars and dust from imaging observations. The first RT modelling of an edge-on galaxy was performed in the optical range, for NGC~891, by \cite{Kylafis_Bahcall1987}. This was the first time a RT model was applied in conjunction with realistic distributions of stars and dust (finite exponential discs as opposed to  slabs or sandwich geometries - \citealp{Disney1989} or spherical configurations - \citealp{Witt1992,WittGordon1996,Witt2000}). Fundamental work on characterising the relative distributions of stars and dust and their statistical trends were later established in a series of papers dedicated to the modelling of edge-on galaxies in the optical and NIR by \cite{Xilouris1997,Xilouris1998,Xilouris1999}.  

The first panchromatic (UV/optical/FIR/submm) RT modelling of an edge-on galaxy was achieved for the same edge-on spiral NGC~891 (see above for the first RT modelling) by \cite{Popescu2000}. Further work on modelling edge-on spirals was performed by \cite{Misiriotis2001,Popescu2004,Bianchi2008,Baes2010,Popescu2011,MacLachlan2011,Robitaille2012,DeGeyter2015,Mosenkov2016,Mosenkov2018,Popescu2017,Natale2021}.

More recently, non edge-on galaxies have also started to be modelled with RT methods. There are two approaches to this type of modelling. The first approach is to use non axi-symmetric RT models: \cite{DeLooze2014,Viaene2017,Williams2019,Verstocken2020,Nersesian2020a,Nersesian2020b}.
Non axi-symmetric models are, in principle,  ideal to capture the detailed geometry seen face-on. However, due to computational limitations, they are implemented to only fit the integrated SED, with typically only four free parameters that describe the total luminosity output for stars and the dust mass, with the geometry being fixed. The second approach is to use axi-symmetric RT models (\citealp{Thirlwall2020}; \citealp{Rushton2022}) and fit not only the integrated SED, but also the geometry of stars and dust, albeit with the assumption of axi-symmetry.
So far only non interacting face-on spiral galaxies have been modelled with axi-symmetric models (M33;~\citealp{Thirlwall2020}, NGC~628;~\citealp{Rushton2022}). The goal of this paper is to extend the applicability of the model to a non-isolated spiral. We will apply the model to M51, since, despite its interaction with a close companion, most of the extent of the galaxy has regular spiral structure, and is not perturbed by the interaction, thus making it suitable for the axi-symmetry approximation.

First discovered by Charles Messier in 1773, M51 is a system of interacting galaxies consisting of M51a (=NGC 5194), the \lq Whirlpool Galaxy\rq\,, a grand-design spiral of Hubble type Sbc, and the smaller lenticular galaxy M51b (=NGC 5195), of Hubble type SB0.
While interaction between the two galaxies is evident from the \lq\lq bridge\rq\rq\, linking them, there is still debate of whether the galaxies have interacted on a single fly-by collision (\citealp{Toomre1972}; \citealp{Durrell2003}) or by multiple close passes (\citealp{Salo2000}; \citealp{Dobbs2010}).
Despite the interaction, as mentioned before, M51a exhibits a good deal of azimuthal symmetry.
This, in conjunction with the close proximity ($\text{D}=8.58\,\text{Mpc}$; \citealp{McQuinn2016}) and the wealth of detailed observational data, makes M51a ideal for studying the energy balance between absorption by dust and re-emission in the far-infrared for a non-isolated spiral galaxy that is seen not far from face-on. In the following we will refer to M51a as M51.

In this work we will utilise the axisymmetric RT models  developed in \cite{Popescu2000}, \cite{Tuffs2004} and  
\cite{Popescu2011}, which have been  successfully used to account for the properties of star forming galaxies derived from statistical samples (e.g. \citealp{vanderGiessen2022}), as well as for the detailed spatial and spectral energy distribution of individual nearby galaxies (\citealp{Popescu2000}; \citealp{Thirlwall2020}; \citealp{Rushton2022}) and of the Milky Way (\citealp{Popescu2017}; \citealp{Niederwanger2019}; \citealp{Natale2021}).
These models describe the geometries of the stellar emissivity and dust opacity from the UV to submm using parameterised analytic functions.
The models consider the absorption and anisotropic scattering of stellar photons with dust grains of various sizes and chemical compositions, with the optical constants taken from \cite{Draine1984} and \cite{Draine2007}. The models also consider the heating of the dust grains and the re-emission in the infrared, including explicit calculations for stochastic heating.

Previous work on modelling M51 was done by \cite{DeLooze2014} and \cite{Nersesian2020b} using non-axisymmetric RT models. As discussed above, these models have the inherent limitation that they cannot self-consistently derive the geometry of emitters and absorbers.  In this study we aim to model M51, by self-consistently deriving the geometry of stars and dust through fitting  the detailed azimuthally averaged UV/optical/FIR/submm surface brightness profiles of M51. As we will show here, our axi-symmetric model produces a good fit to the imaging data within a radial distance of 7 kpc from the centre of M51, within which the galaxy preserves azimuthal symmetry.

The paper is organised as follows: In Sect.~\ref{section:data}
we present the imaging observations used for constraining the model, and the data reduction steps undertaken in this study. In Sect.~\ref{section:AxisymmetricModel} we briefly describe the model and its geometrical components, with the fitting procedure detailed in Sect.~\ref{section:fitting}. In Sect.~\ref{section:results} we present the results of the fits for the surface-brightness distributions and for the intrinsic parameters of M51, including the SFR, dust optical depth,  dust mass, the fractional contribution of the different stellar populations in heating the dust and the attenuation curve. We also present the results for the derived morphological components of M51. In Sect.~\ref{section:discussion} we compare our model for M51 with previous RT modelling of this galaxy and discuss an alternative to the standard model presented in this paper.
We summarise our main results in Sect.~\ref{sec:summary}.

\section{Observational Data}
\label{section:data}
In this section we discuss the data set of M51 used in this study. 
To start with, basic parameters like inclination and position angle were difficult to determine with standard methods, and, because of this, there has been a lot of uncertainty in the value of these parameters.
Determinations of the inclination angle vary from $15^{\circ}$ to $42^{\circ}$ \citep{Danver1942,Toomre1972,Tully1974,Bersier1994,Monnet1981,Considere1982, Shetty2007, Hu2013}, while determinations for the PA vary from $-10^{\circ}$ to $36^{\circ}$ \citep{Burbidge1964, Tully1974, Monnet1981, Considere1982,Shetty2007, Tamburro2008, Walter2008,Hu2013}. We adopted the most recent values from \cite{Hu2013}. The distance, inclination, position angle and central coordinates used in this paper are listed in Table~\ref{tab:distances}.
We obtained panchromatic images from various missions, as summarised in Table~\ref{tab:observations}.

\begin{table}
\caption{Distance, inclination, position    angle and coordinates adopted for the modelling of M51 (M51/NGC~5194), together with the relevant reference.}
\label{tab:distances}
\begin{tabular}{lll}
\hline
\hline
Distance & $8.58\pm0.1$\,Mpc &         \cite{McQuinn2016} \\
Inclination & $20.3^{\circ} \pm 2.8^{\circ}$ & \cite{Hu2013}\\
Position Angle & $12.0^{\circ} \pm 2.5^{\circ}$ & \cite{Hu2013}\\
Right Ascension & $13^{\text{h}}\,29^{\text{m}}\,57.11^{\text{s}}$  &      \cite{Turner-and-Ho1994}\\
Declination & $47^{\circ}\,11^{\prime}\,\,\,42.62^{\prime\prime}$ &     \cite{Turner-and-Ho1994} \\
\hline
\end{tabular}
\end{table}

All data was found on NASA/IPAC Extragalactic Database (NED)\footnote{https://ned.ipac.caltech.edu}, except for the GALEX data, which was obtained from the Mikulski Archive for Space Telescopes (MAST)\footnote{https://archive.stsci.edu/prepds/gcat/}.
Before analysis, the units of the surface brightness maps were converted into [MJy\,sr$^{-1}$]. 

\begin{table*}
    \caption{A summary of the observational data used in this study. The following references were used: (1) \protect\cite{GilDePaz2007}, (2) \protect\cite{Brown2014}, (3) \protect\cite{Bendo2011}.}
    \label{tab:observations}
    \centering
    \begin{tabular}{p{0.16\linewidth}p{0.12\linewidth}p{0.1\linewidth}p{0.1\linewidth}p{0.09\linewidth}p{0.1\linewidth}p{0.13\linewidth}}
    \hline
    Telescope   & Filter/    & $\lambda_{0}$\,[$\mu$m]  & Pixel scale & $\dfrac{\epsilon_{\text{cal}}}{F_\nu}$[\%]   & $\sigma_{\text{bg}}$ $\left[\rm{kJy\, sr^{-1}}\right]$ & Band name \\
     & Instrument & & [$^{\prime\prime}$] & & & \\
    \hline
    GALEX$^1$       & FUV   & 0.1542    & 1.5   & 10  & 0.0019 & FUV \\
    GALEX$^1$       & NUV   & 0.2274    & 1.5   & 10  & 0.0024 & NUV \\
    \hline
    SDSS$^2$        & u     & 0.3562    & 0.396     & 2  & 0.049 & u \\
    SDSS$^2$        & g     & 0.4719    & 0.396     & 2  & 0.017 & g \\
    SDSS$^2$        & r     & 0.6185    & 0.396     & 2  & 0.028& r \\
    SDSS$^2$        & i     & 0.7500    & 0.396     & 2  & 0.049 & i \\
    \hline
    2MASS$^2$       & J     & 1.2000    & 1.0  & 3  & 0 & J \\
    2MASS$^2$       & K$_{\text{s}}$     & 2.2000    & 1.0  & 3  & 0 & K$_{\text{s}}$ \\
    \hline
    Spitzer$^2$     & IRAC  & 3.5070    & 0.75  & 3  & 0 & I1 \\
    Spitzer$^2$     & IRAC  & 4.4370    & 0.75  & 3  & 0 & I2 \\
    Spitzer$^2$     & IRAC  & 5.7390    & 0.75  & 3  & 0.27 & I3 \\
    Spitzer$^2$     & IRAC  & 7.9270    & 0.75  & 3  & 0 & I4 \\
    Spitzer$^2$     & MIPS  & 24.000    & 1.5  & 1  & 0 & MIPS24 \\
    \hline
    Herschel$^3$    & PACS  & 70.000    & 1.4  & 10   & 20.0 & PACS70 \\
    Herschel$^3$    & PACS  & 160.00    & 2.85  & 20   & 0 & PACS160 \\
    Herschel$^3$    & SPIRE & 250.00    & 6.0  & 15   & 0 & SPIRE250 \\
    Herschel$^3$    & SPIRE & 350.00    & 8.0  & 15   & 0 & SPIRE350 \\
    Herschel$^3$    & SPIRE & 500.00    & 12.0  & 15   & 0 & SPIRE500 \\
    \hline
\end{tabular}
\end{table*}
    
\subsection{Background removal}
We performed a curve of growth analysis to determine the radius from the centre of M51 at which the emission of the galaxy falls below the background level. 
This was found to be 25\,kpc.
We also visually inspect each map to ensure that the emission associated with M51 does not extend beyond this point.
The outer sampled radius (30\,kpc) was chosen to ensure a large area for measuring the background.
This sampling region extended beyond the limit of some maps, thus the full sampling area could not be used in those cases.
However, even in the most severe cases, this was a negligible portion of the total sampling area.
We then calculate the average background flux by taking the mean surface brightness through 6 annuli of equal width contained in the region $25\,\text{kpc}\leq R \leq30\,\text{kpc}$.
Although M51 is nearly face-on with an inclination of 20.3$^{\circ}$, the interaction with M51b gives rise to some stripped material in the outer disc, making
the outer isophotes more elongated than the inner ones. This change  in isophotal shape was taken into account when defining the background.  Thus, it was found that the best fitting annuli for the background had an inclination of 40$^{\circ}$ (and the same position angle as the galaxy of 12$^{\circ}$).

\subsection{Foreground star removal}
Foreground stars are visible in the UV/optical/NIR wavelengths and can influence the measured values of emission from M51.
It is therefore important to mask these stars.
For this we first produce a median map where each pixel of the median map is equal to the median of the $3\times3$ pixels around it.
A bias map is also produced, that is principally the same as the median map, but takes the median of the $21\times21$ pixels.
This bias map is used as an additional check to mitigate against false positives when masking noisy data.
Pixels with values that exceed both the median and bias maps are then masked.
After this initial pass, we then manually inspect each mask for regions that may have been missed, or unmask false positives, such as HII regions.
HII regions appear as point sources (similarly to foreground stars) and thus are often falsely masked by this automated process.
These HII regions were identified by eye and the catalogue of \cite{Lee_2011}.
We also masked the companion galaxy M51b and it's diffuse emission.

\subsection{Convolutions}
The SDSS, 2MASS and Spitzer data have a higher resolution than that of our model, which would correspond to 50\,pc at the distance of M51.
We therefore degrade the resolution of these data to match the resolution of our model using a two dimensional Gaussian kernel.
The data at all the other wavelengths used in this study have lower resolutions, therefore, for these wavelengths, we degrade the resolution of our model to match the corresponding ones from the data.
For this we used the kernels of \cite{Aniano2011} to convolve the model with the PSF of the observational instrument.

\subsection{GALEX}
The Galaxy Evolution Explorer (GALEX; \citealp{Martin2005}) was a NASA Small Explorer mission that performed an all-sky survey in the far-ultraviolet (FUV $\sim 0.15\,\mu$m)  and near-ultraviolet (NUV $\sim 0.22\,\mu$m).
Both FUV and NUV data are used in this study and were found on the Mikulski Archive for Space Telescopes (MAST) \footnote{https://galex.stsci.edu/GR6/} and made available by \cite{GilDePaz2007}.
The NUV and FUV data has a pixel scale of 1.5$^{\prime\prime}$ (62\,pc) per pixel and a FWHM of 4.2$^{\prime\prime}$ and 5.3$^{\prime\prime}$, respectively.
The flux calibration error is $\frac{\epsilon_{\text{cal}}}{F_\nu}$ = 10\% following \cite{Morrissey2007}. 
        
\subsection{SDSS}
The Sloan Digital Sky Survey (SDSS; \citealp{York2000}) has observed M51 with two sets of CCD arrays.
Each array has an integration time of 51 s \citep{York2000} capturing images through u, g, r, i and z filters.
In this study we use the u, g, r, and i SDSS bands, with images obtained from NED/IPAC and processed by \cite{Brown2014}.
We do not use the z image as this wavelength is not included in our model fitting.
Each band has a pixel scale of 0.396$^{\prime\prime}$ (16\,pc) and FWHM of 1$^{\prime\prime}$ \citep{Gunn2006}.
Each band has a 2\% rms photometric calibration in all bands \citep{Aihara2011}.
The u band required a correction of -0.04 mag \citep{Brown2014}.
        
\subsection{2MASS}
The Two Micron All Sky Survey (2MASS; \citealp{Jarrett2003}) was performed in the NIR bands, J, H and K$_{\text{s}}$, and observed M51 as part of the Large Galaxy Atlas Survey.
We use the 2MASS data in the J and K$_{\text{s}}$ bands provided on NED/IPAC, and processed by \cite{Brown2014}.
We do not use the H band data as this wavelength is not included in our model.
Both J and K$_{\text{s}}$ bands have a pixel scale of 1$^{\prime\prime}$ (41\,pc) and a FWHM of 2.5$^{\prime\prime}$ \citep{Skrutskie2006}.
The calibration errors for J and K$_{\text{s}}$ are 2\%-3\% \citep{Jarrett2003}.
        
\subsection{Spitzer}
\subsubsection{IRAC}
The Spitzer Space Telescope contains the Infrared Array Camera (IRAC). IRAC is a four channel camera that images the 3.6\,$\mu$m, 4.5\,$\mu$m, 5.8\,$\mu$m and 8.0\,$\mu$m bands simultaneously.
In this study we use all four bands. 
The images were found on NED/IPAC and were processed by \cite{Brown2014}.
Each instrument has a pixel size of 0.75$^{\prime\prime}$ (31\,pc). The 3.6\,$\mu$m, 4.5\,$\mu$m, 5.8$\,\mu$m, and 8.0\,$\mu$m data has a FWHM of 1.66$^{\prime\prime}$, 1.72$^{\prime\prime}$, 1.88$^{\prime\prime}$, and 1.98$^{\prime\prime}$, respectively \citep{Fazio2004}.
The flux calibration error for the IRAC data is 3\% \citep{Fazio2004}.
\subsubsection{MIPS}
The Multiband Imaging Photometer for Spitzer (MIPS; \citealp{Rieke2004}) conducted imaging in three wavebands: 24\,$\mu$m, 70\,$\mu$m and 160\,$\mu$m.
In this study we only use the 24\,$\mu$m data, since at 70\,$\mu$m and 160\,$\mu$m  we use the higher quality data from Herschel.
The data was processed by \cite{Brown2014} and made available at NED/IPAC.
MIPS24 has a pixel scale of 1.5$^{\prime\prime}$ (62\,pc) and a FWHM of 6$^{\prime\prime}$ \citep{Rieke2004}.
The calibration error is 1\% \citep{Rieke2004}.
        
\subsection{Herschel}
\subsubsection{PACS}
M51 is observed by the Photo Array Camera \& Spectrometer (PACS; \citealp{Poglitsch2010}) as part of the Very Nearby Galaxy Survey (VNGS) at the 70\,$\mu$m and 160\,$\mu$m bands.
Both bands are utilised in this study and images were obtained from NED/IPAC.
The pixel scale of the 70\,$\mu$m and 160\,$\mu$m data is 1.4$^{\prime\prime}$ (58\,pc) and 2.85$^{\prime\prime}$ (118\,pc), respectively.
The FWHM of the 70\,$\mu$m and 160\,$\mu$m data are 5.6$^{\prime\prime}$ and 11.4$^{\prime\prime}$, respectively \citep{Bendo2011}.
\subsubsection{SPIRE}
The Spectral and Photometric Imaging Receiver (SPIRE; \citealp{Griffin2010})  observed M51 as part of the Very Nearby Galaxy Survey (VNGS) at the 250\,$\mu$m, 350\,$\mu$m and 500\,$\mu$m bands.
All three bands were used in this study and images were obtained from NED/IPAC and processed by \cite{Bendo2011}. The pixel scale of the 250\,$\mu$m, 350\,$\mu$m and 500\,$\mu$m data is 6$^{\prime\prime}$ (250\,pc), 8$^{\prime\prime}$ (333\,pc), and 12$^{\prime\prime}$ (500\,pc), respectively.
The FWHM of SPIRE data for 250\,$\mu$m, 350\,$\mu$m, 500\,$\mu$m are 18.2$^{\prime\prime}$, 24.5$^{\prime\prime}$, and 36$^{\prime\prime}$, respectively \citep{Bendo2011}.
 Flux calibration error is accurate to $\pm 15\%$\citep{Griffin2010}.\\

\begin{figure}
    \centering
    \includegraphics[width=\columnwidth]{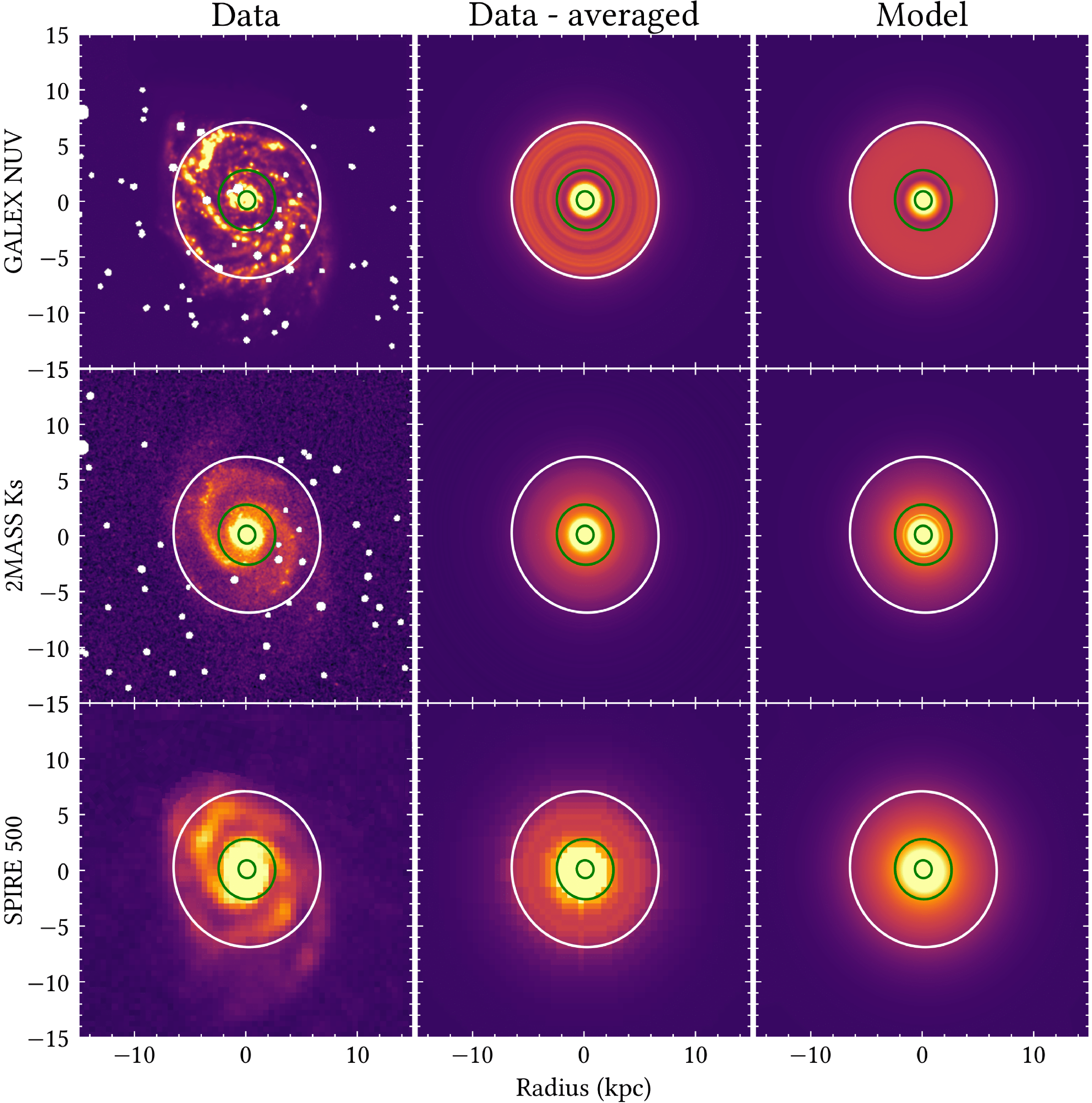}
    \caption{The observed (left), azimuthally averaged observed (middle) and model (right) images of M51 of various wavebands. The emission from M51b is masked out from the figure. We indicate on each image elliptical apertures (green) with radial distances of 0.8 and 2.7\,kpc, 
    which correspond to the inner radius of the inner and main disc of our model, respectively. We also plot in white the elliptical aperture corresponding to $R_l=7$\,kpc, the radius within which the assumption of axi-symmetry holds. The masked foreground stars are indicated by  the white dots.}
    \label{fig:maps}
\end{figure}

\noindent
Examples of  imaging data used in this analysis are given in the left panels of Fig.~\ref{fig:maps}. They show 15x15\,kpc images of M51 in the GALEX NUV, 2MASS K band, and the SPIRE 500\,${\mu}$m.

\subsection{Photometry}

The observed images were corrected for Galactic extinction using a value of $E(B-V) = 0.0308$ \citep{Schlafly2011} and $R_V =3.1$ \citep{Fitzpatrick1999}. Azimuthally averaged surface brightness profiles and spatially integrated flux densities $F_{\nu}$ were derived using curve of growth (CoG)  photometry.  Each map was segmented into 250 annuli of a fixed width in which the average brightness was calculated, with the corresponding radius taken as the midpoint of the annulus.
The elliptical annuli were defined by the inclination and position angle of M51, as  listed in \hyperref[tab:distances]{Table~\ref{tab:distances}}.
Examples of  azimuthally averaged images  (NUV, K, and 500\,${\mu}$m) produced in this way are given in the middle panels of Fig.~\ref{fig:maps}. Comparison between azimuthally averaged and  corresponding original observed images shows that the assumption of axi-symmetry seems to hold up to $\sim 7$\,kpc, which we take to represent the limit of applicability of our axi-symmetric model. We will refer to this limit as $R_l$.

The photometric errors due to background fluctuations, calibration and configuration noise were derived using the procedure from \cite{Thirlwall2020}.

\section{The Model}
\label{section:AxisymmetricModel}
Our model for M51 follows the axisymmetric RT models of \cite{Popescu2011}, hereafter \citetalias{Popescu2011}, which describe the dust opacity and stellar emissivity for spiral galaxies from the UV to submm wavelengths using parameterised analytic functions.
The model considers the absorption and anisotropic scattering of stellar photons with dust grains of various size and chemical composition.
A dust composition consisting of silicates, graphites and PAH molecules was used, with a grain size distribution from \cite{WeingartnerDraine2001}, and optical constants from \cite{Draine1984} and \cite{Draine2007}. As in our previous work on modelling spiral galaxies, we consider a Milky Way-type dust, since this  has proven to provide consistent results for all modelled galaxies. The parameters describing the grain size distribution and grain mixture are taken from Table 1, line 7 of \cite{WeingartnerDraine2001}.
The RT model of M51 also incorporates explicit calculations for the stochastic heating of the dust grains. 

The RT codes used for the modelling of M51 were  the \citetalias{Popescu2011} code and the DART-RAY code \citep{Natale2014,Natale2015,Natale2017}. The \citetalias{Popescu2011} code  is a modified version of the code of  \cite{Kylafis_Bahcall1987}. Both the \citetalias{Popescu2011} and DART-RAY codes use ray-tracing algorithms.

Following \cite{Thirlwall2020}, we adapt the model of \citetalias{Popescu2011} to allow for additional morphological components.
In the case of M51 it was found that, in addition to the bulge, an inner and a main disc component were present out to $R_l$.
Each morphological component comprises the generic stellar and dust components from \citetalias{Popescu2011}: the stellar disc, the thin stellar disc, the dust disc and thin dust disc. The stellar disc is made up of all the stars that are old enough to have had time to migrate and form a thicker configuration (a few hundred parsec scale height), while the thin stellar disc is made up by young stars, that are still spatially closely related to the molecular layer from which they formed (50-100 parsec scale-height). A clumpy component associated with the star forming clouds is also incorporated in the model. All these are described in detail in \citetalias{Popescu2011}. Following \cite{Thirlwall2020} we also found the need to alter the analytic functions describing the exponential discs, to include an inner truncation radius. Thus, we use Eq. (1) and (2) from \cite{Thirlwall2020} to describe
the stellar volume emissivity and the dust density distribution for each disc component j:

\begin{equation}
\label{eq:emissivities}
w_\text{j}(R,z) = 
\begin{cases}\!
  0, &\text{if }R<R_{\text{tin,j}} \\
  \\
  \begin{aligned}[b]
    A_{0,\text{j}}\left[\frac{R}{R_{\text{in,}\text{j}}}\left(1-\chi_{\text{j}}\right)+\chi_{\text{j}} \right]\\
    \times \text{exp}\left(-\frac{R_{\text{in},\text{j}}}{h_{\text{j}}}\right)\text{exp}\left(-\frac{z}{z_{\text{j}}}\right) ,
  \end{aligned} &\text{if}R_{\text{tin,j}}\le R<R_{\text{in,j}}\\
  \\
A_{0,\text{j}}\times
    \text{exp}\left(-\frac{R_{\text{in},\text{j}}}{h_{\text{j}}}\right)\text{exp}\left(-\frac{z}{z_{\text{j}}}\right), & R_{\text{in},\text{j}} \le R \le R_{\text{t},\text{j}}
\end{cases}
\end{equation}
        
with:
\begin{align}
    \label{eq:chi}
    \chi_{\text{j}} = \frac{w_{\text{j}}(0,z)}{w_{\text{j}}(R_{\text{in},\text{j}},z)}
\end{align}
    where $R$ and $z$ are the radial and vertical coordinates, $h_\text{j}$ and $z_\text{j}$ are the scale-length and scale-height respectively, $A_{0,\text{j}}$ is a constant scaling the amplitude, $\chi_\text{j}$ describes a linear slope of the radial distributions from the inner radius, $R_{\rm{in}}$, to the centre of the galaxy.
$R_{\text{tin}}$ and $R_{\text{t}}$ are the inner and outer truncation radii.

Eqs.~\ref{eq:emissivities} and~\ref{eq:chi} are wavelength dependent.
The spatial integration of these formulae (Eqs. B1-B3 from \citealp{Thirlwall2020}) between $R_{\text{tin}}$ and $R_{\text{t}}$ provides the intrinsic stellar luminosity $L_{\nu}$ if j represents a stellar disc component, or the dust mass $M_{\rm d}$ if j represents a dust disc component.

The emissivity of the bulge is described by the  Sérsic distribution

\begin{equation}
\label{eq:bulge}
w_{\text{bulge}}(\lambda,R,z) = w(\lambda,0,0)\, 
\sqrt{\frac{b_{\rm s}}{2\pi}}\,\frac{(a/b)}{R_{\rm e}}\,
\eta^{(1/2n_{\rm s})-1}
\exp{(-b_{\rm s}\, \eta^{1/n_{\rm  s}})}
\end{equation}

with:

\begin{equation}
\label{eq:bulge1}
\eta(\lambda,R,z) = \frac{\sqrt{R^2 + z^2(a/b)^2}}{R_{\rm e}}
\end{equation}
$\eta^{\text{bulge}}(\lambda,0,0)$ is the stellar emissivity at the centre of the bulge, $R_{\text{e}}$ is the effective radius of the bulge, $a$ and $b$ are the semi-major and semi-minor axes of the bulge, respectively, and $b_{\text{s}}$ is a constant which depends on the value of the Sérsic index $n_{\text{s}}$ (see Eq.~8 from \citealp{Natale2021}).
Following \cite{Thirlwall2020}, the parameters of the different components are as follows:\\

\noindent
{\it The stellar disc:}

\noindent
The stellar disc is made up of the old stellar population, defined as a population of stars that had time to migrate from their molecular layer into a thicker configuration (a few hundred parsec scale height). Following from \cite{Popescu2000} and \citetalias{Popescu2011}, we make the approximation that the stellar disc only emits in the optical/NIR (no UV counterpart).
It is described by the four geometrical parameters 
$h_{\rm{s}}^\text{disc}$, $z_{\rm{s}}^\text{disc}$, $R_{\rm{in,s}}^\text{disc}$ and $\chi_{\rm{s}}^\text{disc}$, and one amplitude parameter $L^\text{disc}$. The free parameters are 
$h_{\rm{s}}^\text{disc}$, $\chi_{\rm{s}}^\text{disc}$ and $L^\text{disc}(\lambda)$. $z_{\rm{s}}^\text{disc}$ is fixed from the model, and $R_{\rm{in,s}}^\text{disc}$ is fixed from the data. We allow $L^{\text{disc}}$ to be a free parameter for each wavelength corresponding to u, g, r, i, J, K, I1, I2, I3.
We also allow $h_{\rm{s}}$ to be wavelength dependent, hence we consider the scale-lengths for the g, r, i, J, K, I1, I2, I3 bands to be free parameters. The scale-length in the u-band is taken to be the same as in the g-band. The model of M51 includes an inner and a main stellar disc.\\

\noindent
{\it The thin stellar disc}

\noindent
The thin stellar disc is made up of the young stellar population of stars that did not have time to migrate from their molecular layer (50-100 pc). This disc emits predominantly in the UV, although we assume a smooth decrease of its emission in the optical/NIR.
It is described by the four geometrical parameters $h_{\rm{s}}^\text{tdisc}$, $z_{\rm{s}}^\text{tdisc}$, $R_{\rm{in},s}^\text{tdisc}$ and $\chi_{\rm{s}}^\text{tdisc}$, and one amplitude parameter $L^\text{tdisc}$.  The free parameters are $h_{\rm{s}}^\text{tdisc}$, $\chi_{\rm{s}}^\text{tdisc}$ and $L^\text{tdisc}$. $z_{\rm{s}}^\text{tdisc}$ is fixed from the model and $R_{\rm{in},s}^\text{tdisc}$ is fixed from the data. The scale-length of this disc is assumed to be wavelength independent.  The wavelength dependence of $L^\text{tdisc}$ is fixed from the model. The model of M51 includes an inner and a main thin stellar disc.

Keeping in line with previous modelling, we express the spectral integrated luminosity of the young stellar disc $L^{\rm tdisc}$ in terms of a star formation rate ${\rm SFR}$, using equations (16), (17), and (18) from \citetalias{Popescu2011}.\\

\noindent
{\it The dust disc}

\noindent
It is described by the geometrical parameters $h^{\rm{disc}}_{\rm{d}}$, $z_{\rm{d}}^{\rm{disc}}$, $R_{\rm{in,d}}^{\rm{disc}}$ and $\chi_{\rm{d}}^{\rm{disc}}$ and the amplitude parameter, the face-on optical depth of the B-band at the inner radius $\tau_{\rm{B}}^{\rm{f, disc}}\left(R_{\rm{in,d}}^{\rm{disc}}\right)$. The free parameters are $h^{\rm{disc}}_{\rm{d}}$, $\chi_{\rm{d}}^{\rm{disc}}$ and $\tau_{\rm{B}}^{\rm{f,disc}}\left(R_{\rm{in,d}}^{\rm{disc}}\right)$. $z_{\rm{d}}^{\rm{disc}}$ is fixed from the model and $R_{\rm{in,d}}^{\rm{disc}}$ is fixed from the data. Our model of M51 contains an inner, and main  dust disc.\\

\noindent
{\it The thin dust disc}

\noindent
The thin dust disc is described by the geometrical parameters $h^{\rm{tdisc}}_{\rm{d}}$, $z_{\rm{d}}^{\rm{tdisc}}$, $R_{\rm{in,d}}^{\rm{tdisc}}$ and $\chi_{\rm{d}}^{\rm{tdisc}}$ and the amplitude parameter, the face-on optical depth of the B-band at the inner radius $\tau_{\rm{B}}^{\rm{f, tdisc}}\left(R_{\rm{in,d}}^{\rm{tdisc}}\right)$.
Following \citetalias{Popescu2011} we fix the geometric parameters  of the thin dust disc to that of the thin stellar disc. The model of M51 contains an inner, and main thin dust disc.\\

\noindent
{\it The clumpy component}

\noindent
The clumpy component represents the  dense dusty clouds associated with star forming regions that are illuminated by the strong, UV-dominated radiation fields produced by the photons emitted from the young stars inside the clouds.
The dust in these regions reaches thermal equilibrium with the intense radiation fields and emits dominantly in the 24\,$\mu$m to 70\,$\mu$m region, often surpassing the emission due to the diffuse component at these wavelengths.
Following \citetalias{Popescu2011} we define a \lq\lq clumpiness factor\rq\rq\, $F$, which is the amplitude parameter of the clumpy component, and represents the fraction of the total luminosity of massive stars that is locally absorbed by the dust in these star-forming clouds. The diffuse component is then illuminated by the escape fraction $1-F$. The $F$ factor has a wavelength dependence that is not determined by the optical properties of the dust grains, but by the evolutionary stage of the birth clouds. This is because in our model the dust cloud is very opaque, thus completely blocking any radiation from inside the cloud. The only way photons escape is through holes in the cloud, produced by the fragmentation in the cloud.  The wavelength dependence of the escaping radiation thus arises because
lower mass and less blue stars spend a higher proportion of their
lifetime radiating when they are further away from their birth-clouds, and because of the progressive fragmentation of the clouds. The formalism is described in \cite{Tuffs2004}. The dust emission SED of the birth clouds is taken to follow the prescription from \citetalias{Popescu2011}.

\section{Fitting the model}
\label{section:fitting}
\begin{figure*}
    \centering
    \includegraphics[scale=6.82]{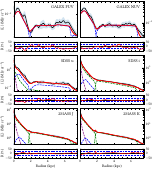}
    \caption{Comparison between the azimuthally averaged surface brightness profiles of the observations (solid black line) at selected UV/optical/NIR wavelengths and of the corresponding model for dust attenuated stellar light (solid red line). 
    The blue shaded region around the observed profile represents the corresponding errors in the averaged surface brightness, as described in Sect.~\ref{section:data}.
    The contribution from each morphological component is plotted with dashed lines and  colour-coded as follows: bulge in purple, stellar disc in green and thin stellar disc in blue.
    The lower panels show the residuals between the observations and our model, with the dashed blue lines showing the $\pm20\%$ residuals to guide the eye.}
    \label{fig:wd01_uv}
\end{figure*}

\begin{figure*}
    \centering
    \includegraphics[scale=6.82]{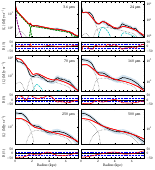}
    \caption{Same as Fig.~\ref{fig:wd01_uv} but for selected wavelengths in the NIR/MIR/FIR/submm, where the emission from the dust discs is plotted with the grey dotted lines and the dust emission from the HII regions is shown by the dashed cyan lines.}
    \label{fig:wd01_dust}
\end{figure*}

To directly compare the model to the data we produce azimuthally averaged surface brightness profiles of both. The observed surface-brightness profiles were obtained by carrying out curve of growth photometry on the maps listed in \hyperref[tab:observations]{Table~\ref{tab:observations}}, as described in Sect.~\ref{section:data}. For the model profiles we carried out an identical procedure.
Examples of these profiles can be seen in Figs.~\ref{fig:wd01_uv}-\ref{fig:wd01_dust}. Examples of model images are given in the right panels of Fig.~\ref{fig:maps}.

From an initial inspection of the surface brightness profiles of the observational data it was clear that a single disc described by {Eqs.~\ref{eq:emissivities}-\ref{eq:chi}} would provide an inadequate fit.
Following \cite{Thirlwall2020}, we segment the galaxy into distinct morphological components to fully capture the detailed morphology of M51.
As mentioned in Sect.~\ref{section:AxisymmetricModel}, we found that, within $R_l$, in addition to the bulge, two distinct morphological components were needed to model this galaxy: an inner and a main disc. 
We allow for each morphological component to be made up of two stellar and two dust components.

We followed the same fitting procedure as in \cite{Thirlwall2020}. In brief, this involves an intelligent searching algorithm of the parameter space, that takes into account the fact that different parameters do not equally affect the emission at all wavelengths. This allows one to identify pairs of parameters that predominantly shape emission at key wavelengths and use this to fit one pair (usually made of a geometric parameter and an amplitude parameter) of parameters at a time.

Thus, we begin modelling with some initial guess parameters.
These parameters are based on notable features of the surface brightness profiles, as well as informed by previous modelling \citep{Thirlwall2020}.
We start by fitting the NUV profile. This is equivalent to constraining the spatial distribution of the young stellar population residing in the thin stellar disc.
Although this stellar disc is highly attenuated in the UV, we can still derive an estimate of the corresponding geometrical and amplitude parameters for an initial guess of dust opacity. We thus derive an exponential scale length $h^{\rm{tdisc}}_{\rm{s}}$ and a luminosity density $L^{\rm{tdisc}}$ in the NUV filter, for the inner and main disc, assuming a known wavelength dependence of the luminosity density at the other sampled wavelengths. There is also the free parameter $\chi_{\rm s}^{\rm tdisc}$, which is fitted for the inner disc in order to account for the fall of emissivity towards the galaxian centre. For the main disc $\chi_{\rm s}^{\rm tdisc}$ is only used to provide a smooth overlap between the inner and the main disc,  resembling the observed averaged profiles, instead of simply using delta functions. $\chi_{\rm s}^{\rm m-tdisc}$ is therefore not a critical parameter in the fit. Inner radii $R_{\rm in}$ and inner and outer truncation radii $R_{\rm tin}$ and $R_{\rm t}$ are fixed from data and are not subject to iterations.
Once we obtain a good fit in the NUV (for the trial dust opacity), we then attempt to constrain the dust distribution .
To do this we re-run the RT calculation with the parameters derived from fitting the NUV profile and for an initial guess for the optical emission, and predict the dust emission at all infrared wavelengths, including the 500\,$\mu$m.  We then constrain the spatial distribution of the dust by comparing the predicted 500\,$\mu$m model profiles with the corresponding observed ones, since at this wavelength the dust emission primarily traces dust column density rather than heating sources. At this step small changes in the dust parameters may be needed for a best fit to the 500\,$\mu$m data.  This results in a fit for the scale-length of the dust disc $h^{\rm{disc}}_{\rm{d}}$ and a dust opacity (plus the parameter $\chi_{\rm d}^{\rm tdisc}$). A few iterations are needed between NUV and the 500\,$\mu$m,  until a good fit at both NUV and the 500\,$\mu$m is achieved. Towards the end of the iteration the FUV luminosity density is also slightly adjusted until the best fit to the FUV profile is also obtained. Once the parameters related to the thin stellar disc and dust disc are constrained,  
we run again the RT code to derive the distribution of the stellar disc and bulge harbouring the old stellar population, by fitting the optical and NIR profiles at each wavelength for which observational data is available. Since in a face-on galaxy the optical and NIR emission is fairly optically thin to dust, the final fits at these wavelengths do not alter the solution for for the dust disc and the thin stellar disc (in the UV) obtained in previous iterations. A clumpy component is finally fitted to account for any remaining 24-70\,${\mu}$m observed emission not accounted for by the diffuse dust emission. Overall, within a small number of iterations a solution is derived for the whole spectral range. Notably, the 70-350\,${\mu}$m emission is not directly used in the fit, but instead is predicted. To conclude, the fit mainly consists in fitting only a pair of parameters (a scale-length and an amplitude) at any given time, for each morphological component at a given wavelength.
More details on the fitting procedure can be found in \cite{Thirlwall2020}.

The main (free) geometric parameters constrained by the procedure outlined above are listed in Table~\ref{tab:geometry_combined}. The free parameters related to stellar luminosity densities are listed in Table~\ref
{tab:int_lum_wd01}. The free parameters related to dust opacity are listed in Table~\ref{tab:optical depth}. Other parameters of the model are listed in Table~\ref{tab:geometry regions}.

\begin{table*}
\centering
\begin{tabular}{l|l}
    \hline
    $h_{\text{s}}^{\text{i-disc}}$\text{(g, r, i J, K, I1, I2, I3)} & (0.74, 0.73, 0.72, 0.69, 0.66, 0.70, 0.72, 1.0)$\pm$7\% \\
    $h_{\text{s}}^{\text{m-disc}}$\text{(g, r, i J, K, I1, I2, I3)} & (3.8, 3.5, 3.7, 3.5, 3.5, 4.3, 4.9, 7.0)$\pm$13\% \\
    \\
    $h_{\text{s}}^{\text{i-tdisc}}$ & 0.6$\pm$0.15 \\
    $h_{\text{s}}^{\text{m-tdisc}}$ & 4.3$\pm$1.0 \\
    \\
    $h_{\text{d}}^{\text{i-disc}}$ & 5.0$^{+3.5}_{-1.5}$ \\
    $h_{\text{d}}^{\text{m-disc}}$ & 6.0$^{+3.0}_{-0.9}$ \\
    \\
    $\chi_{\text{s}}^{\text{i-tdisc}}$ &0.1$\pm$0.02 \\
    $\chi_{\text{s}}^{\text{i-disc}}$ & 0 \\
    $\chi_{\text{d}}^{\text{i-disc}}$ & 0.45$\pm0.05$  \\
    \\
    $R_\text{eff}$ & 0.35$\pm$0.1 \\
    \hline
\end{tabular}
\caption{Main geometrical (free) parameters of the inner and main discs, and of the bulge, derived for the standard model.}
\label{tab:geometry_combined}
\end{table*}
The results of the fits are presented in Figs.~\ref{fig:wd01_uv}-\ref{fig:wd01_dust}, showing an overall reasonable agreement with the data. To help the comparison, the residuals $R$ between the surface brightness profiles of the observations and of the model are also plotted on the bottom panels of Figs.~\ref{fig:wd01_uv}-\ref{fig:wd01_dust}. The residuals are defined as:
\begin{equation}
    R = \frac{\text{model}-\text{observation}}{\text{observation}}
\end{equation}
In addition, the goodness of fit is also considered with the reduced chi-squared  $\text{chi}^{2}_{r}$ statistics\footnote{We use the notation $\text{chi}^{2}$ rather than $\chi^{2}$, as the later notation is reserved in this work for the geometrical parameter describing the inner linear slope.}, which becomes more relevant when finalising the model solution.
We calculated $\text{chi}^{2}_{r}$ for each wavelength by the following formulae:

\begin{align}
    & \text{chi}^{2}_{\lambda}=\sum_{n=1}^{N}\frac{\left(M_n-O_n\right)^2}{\epsilon^{2}_{\text{SB,}n}} \\
    & \text{chi}^{2}_{\text{r,}\lambda}=\frac{\text{chi}^{2}_{\lambda}}{N-m}\approx \frac{\text{chi}^{2}_{\lambda}}{N}
    \label{eq:reduced_chi_sqr}
\end{align}
where $N$ is the total number of annuli $n$, $m$ is the number of free parameters fitted at wavelength ${\lambda}$, $O_{\text{n}}$ is the azimuthally averaged flux through annuli $n$, $M_{\text{n}}$ is the model flux through annuli $n$, $\epsilon^{2}_{\text{SB,}n}$ is the total error of the azimuthally averaged surface brightness profile corresponding to annuli $n$. The total number of annuli is 70 while $m$ varies between 2 in the FUV, to 5 (in the NUV and 500\,${\mu}$m), to 8 (in the optical and NIR) for a fitted profile out to 7 kpc radius. At most other wavelengths that are predicted, we can assume zero free parameters. Because $N$ is much larger than $m$ we neglect $m$ in Eq. \ref{eq:reduced_chi_sqr}.
Values of the $\text{chi}^{2}_{\lambda}$, for selected wavelengths, are listed in Table~\ref{tab:chi_sqr_wd01}. The total reduced chi-squared statistic over all wavelengths is:
\begin{equation}
    \text{chi}_{\rm{r}}^{2}=\frac{\sum_{l=1}^{L}\text{\text{chi}}^{2}_{\lambda, l}}{\sum_{l=1}^{L}N_{l}}
\end{equation}

\begin{table}
    \centering
    \begin{tabular}{c|c}
    \hline
        Band & $\text{chi}^{2}_{r}$ \\
        \hline
        NUV & 1.09 \\
        I2 & 0.74 \\
        PACS70 & 3.62 \\
        SPIRE500 & 1.33\\
        \hline
        Global & 2.44
        
    \end{tabular}
    \caption{The $\text{chi}^{2}_{r}$ values at selected wavelengths, as well as the global $\text{chi}^{2}_{r}$ (over all wavelengths), for the standard model.}
    \label{tab:chi_sqr_wd01}
\end{table}

 The  $\text{chi}^{2}_{r}$  values from Table~\ref{tab:chi_sqr_wd01} confirm the trends seen in Figs.~\ref{fig:wd01_uv}-\ref{fig:wd01_dust}, proving that the axi-symmetric model provides a reasonable fit to the data.

\section{Results}
\label{section:results}

\subsection{The surface brightness distributions}
Taking the azimuthally averaged surface brightness of the observations produces relatively smooth exponential profiles, as seen in Figs.~\ref{fig:wd01_uv}-\ref{fig:wd01_dust}.
Despite the slight imperfections in the smoothness, the average profiles are still  suitable for description with analytic axi-symmetric functions.
The most significant deviations are in the UV and 24$\,\mu$m bands, where strong asymmetries and clumpiness are present.
The bumps in these profiles generally coincide with the spiral arms as well as star forming regions.
For example, a significant bump in emission at 1.3\,kpc, which can be seen most notably in the FUV profile, is due to bright star forming regions.
There is also a bump in emission between 5\,kpc and 6.5\,kpc, which coincides with an overlap in the spiral arms.
Predictably, the 24\,$\mu$m profiles exhibit the least smooth curves, as at this wavelength the emission due to the diffuse dust and the dust heated by localised SF regions is roughly equal, hence strong asymmetries  arise from these localised sources of emission.
This is also  seen in the IRAC bands, where there is significant PAH emission.

Overall the model has reasonable agreement with the data with residuals generally within 20\%.

The scale-length of the observed profiles in the optical regime generally decreases with increasing wavelength for each morphological component.
For wavelengths $\geq3.6\,\mu$m this trend reverses for the main disc. 
This is similar with trends found from previous modelling (M33; \citealp{Thirlwall2020}).

The observed profiles in the IRAC bands are modelled with emission from the old stellar populations as well as dust emission, with a negligible contribution from the young stellar population.
At $3.6\,\mu$m and $4.5\,\mu$m the emission is dominated by the old stellar population, but at $5.8\,\mu$m the dust emission is almost equal in amplitude to that of the old stellar population.
At the $8.0\,\mu$m band the emission is dominated by the dust, primarily due to the PAH emission.

\subsection{The global SED of M 51}
\label{section:global-sed}
\begin{figure}
    \centering
    \includegraphics[width=\columnwidth]{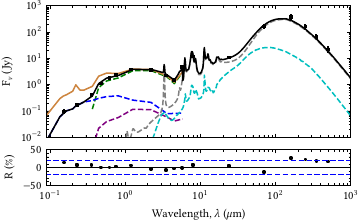}
    \caption{The global SED of the model (spatially integrated out to $R_{\rm l}$) plotted together with the observed flux densities 
    used to optimise our model (shown with black squares).
    The error bars of the data are also shown, although most are contained within the square symbols.
    The model SED is represented by the solid black line, and covers both the dust attenuated stellar SED in the UV/optical/NIR as well as the dust emission SED in the NIR/MIR/submm. The contributions from each component of the model are plotted with dashed lines as follows: thin stellar disc in dark blue, stellar disc in green, bulge in purple, diffuse dust in grey and clumpy component in cyan.
    The intrinsic stellar SED (as would be seen in the absence of dust) is plotted with the solid brown line.
    We plot the relative residuals in the lower panel, with the $\pm20\%$ residuals indicated with the dotted blue line to guide the eye.}
    \label{fig:wd01-sed-all}
\end{figure}

\begin{figure}
    \centering
    \includegraphics[width=\columnwidth]{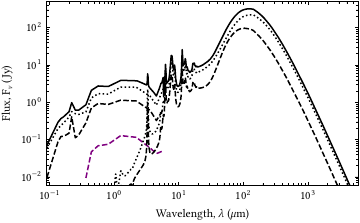}
    \caption{The predicted global SED of M 51a (solid black line) for the intrinsic (dust de-attenuated stellar emission) and for the dust emission, plotted together with the contribution from the individual morphological components: the inner (dashed black line) and  the main (dotted black line) discs, and the bulge (dashed purple line).}
    \label{fig:wd01-sed-components}
\end{figure}

The global SED of M51 was produced by spatially integrating the emission at each wavelength out to ${\rm R}_l$. This was done both for observations and for the model and the results are plotted in Fig.~\ref{fig:wd01-sed-all}.
The  model SED closely resembles the observed SED, with a maximum residual of $|R|=26.2\%$ and an average residual of $\langle|R|\rangle=4.33\%$. The model performs worst at $160\,\mu$m, where the emission is underestimated by our model.

We also show the contributions from the different morphological components to the global SED in Fig.~\ref{fig:wd01-sed-components}. 
It is important to note that the dust emission for a given morphological component is not entirely powered by the stellar emission of the same component, as for example, some fraction of the stellar photons emitted from the inner disc will contribute to heating the dust in the main disc. Nonetheless, we still show the separate contributions of the inner and main discs to the dust emission, within the accepted understanding that there is not a one-to-one spatial correspondence between the stellar and the dust emission of these components.
As expected due to its extent, the largest contribution to the global SED is from the main disc responsible for 67.4\% of the total stellar emission, and 63.7\% of the dust emission.
The inner disc contributes 30.7\% to the total stellar emission and 36.3\% to total dust emission.
The bulge contributes 1.89\% to the total stellar emission. Here we note that the separation between the inner disc and the bulge is rather degenerate. We found no strong constraints for this separation. We therefore consider the bulge  versus inner disc luminosity to be subject to large uncertainties.

The dust emission SED is dominated by the diffuse component. The HII component has a smaller contribution, of $12.2\%$.  We find a clumpiness factor of $F=0.125$. The global dust emission SED peaks at $130\,\mu$m, which corresponds to a dust temperature of 22\,K. The hottest dust resides within the inner disc, where the SED peaks at $102\,\mu$m, corresponding to a dust temperature of 28\,K. The main disc has the greatest contribution to the global SED, therefore, as expected, it peaks at a similar wavelength, at around $133\,\mu$m, corresponding to a dust temperature of 22\,K.

We find that 38.3$\pm4$\% of the stellar light is absorbed and re-radiated by the dust.
This is a typical value for late-type spiral galaxies \citep{Popescu2002}.
M33 was found to have a similar value of 35$\pm$3\% \citep{Thirlwall2020}.
We also find that the young stellar population contributes 42.8\% to the total intrinsic stellar light, with 55.4\% being due to the old stellar population {in the disc} and the remaining 1.89\% due to the bulge.

\subsection{Star-formation rates in M51}
We obtain a $\text{SFR}=4.11^{+0.41}_{-0.39}\,$\mpyr\ which is within the broad range of literature values.
\cite{DeLooze2014} obtained a $\text{SFR}\sim3.13 \,$\mpyr(\footnote{Value adjusted for the distance of $D=8.58$\,Mpc, as used in this study.}) and \cite{Nersesian2020b} found a $\text{SFR}=4.09\pm0.35\,$\mpyr(\footnote{Adjusted value.}). Both  these studies use SKIRT, a Monte-Carlo RT model.
\cite{Calzetti2005} obtain the estimates $\text{SFR}(24\,\mu\text{m})\sim3.7\,$\mpyr(\footnote{Adjusted value.}) and $\text{SFR(FUV)}\sim4.7\,$\mpyr(\footnote{Adjusted value.}).

About two thirds of the star formation takes place in the main disc with $\text{SFR}^{\text{m}}=2.73^{+0.28}_{-0.26}\,$\mpyr.
This is expected due to both the size and bolometric output of the main disc.
For the inner disc we find $\text{SFR}^{\text{i}}=1.38^{+0.15}_{-0.14}\,$\mpyr.
We also calculate the surface density of star formation rate, $\Sigma_{\text{SFR}}$, for each morphological component, by considering the physical extent of each disc, which is shown in Tables~\ref{tab:sfr}-\ref{tab:surface_sfr}. The global surface density in SFR is $\Sigma_{\text{SFR}}=2.83^{+0.28}_{-0.27}\times10^{-2}\,$\mpyrkpc.

\subsection{Dust optical depth and dust mass}
We find that the face-on optical depth in the B-band has a maximum value of $\tau_{\text{B}}^{\text{f}}\left(R_{\text{in}}\right)=5.33\pm0.33$ at the inner radius of the inner disc.
The main disc has a face-on optical depth at the inner radius of $2.62\pm0.08$.
These values are listed in Table~\ref{tab:optical depth}.
We find that M51 has a global dust mass (out to $R_l$) of M$_{\text{d}}=3.74^{+0.34}_{-0.30}\times10^{7}\,\text{M}_{\odot}$.
The dust masses of the individual morphological components are listed in Table~\ref{tab:dust_mass}.

Using a gas mass of M$_{\text{G}}=1.1\pm0.03\times10^{10}\,\text{M}_{\odot}$ \citep{Mentuch_cooper_2012} we obtain a gas-to-dust ratio of $\text{GDR}=294\pm4$.

\subsection{Morphological components of M51}
In modelling M51 we found three distinct morphological components out to  $R_l$, each requiring different geometrical parameters for the distribution of stars and dust.
Although we present the model for the entire galaxy, it is also interesting to study the properties of these individual components.
The geometrical parameters of the three morphological components, as constrained by our model are listed in Table~\ref{tab:geometry_combined}.

\subsubsection{The inner disc}
The inner disc extends from 0\,kpc to 1.9\,kpc and exists as a  ring-like structure for the young stars and dust, with an inner radius of 700\,pc and 800\,pc respectively.
The distribution of the old stars, however, exponentially increases towards the centre.
The distribution of the dust is notably flat with a scale length of 5\,kpc, compared to a scale length of 600\,pc for the young stellar distribution.
The inner disc has a $\text{SFR}^{\text{i}}=1.38^{+0.15}_{-0.14}\, $\mpyr, contributing around 30\% to the global SFR, and has the greatest surface density of star formation rate  $\Sigma_{\text{SFR}}^{\text{i}}=1.21^{+0.28}_{-0.11}\times10^{-1}\,$\mpyrkpc.
The dust mass within the inner disc is M$_{\text{d}}^{\text{i}}=6.39^{+0.37}_{-0.21}\times10^{6}\,\text{M}_{\odot}$.
The temperature of the dust contained within the inner disc is the hottest in the galaxy, at around 28\,K.

\subsubsection{The bulge}
The bulge was found to have an effective radius of 350\,pc and was modelled with a Sérsic index of 4.
As mentioned in Sect.~\ref{section:global-sed}, the bulge parameters are subject to uncertainty, in the sense that there are different configurations in the bulge/inner disc decomposition that could result in a good fit to the data. We based our choice on covering a small dip in the profile, that would otherwise appear in other configurations, and as such we believe that our choice is the most reliable result. Nonetheless, we caution the reader of the subjectivity of the approach. The error estimates on the bulge/inner disc parameters do not take into account the uncertainty due to the other possible combinations.

\subsubsection{The main disc}
The main disc resides in the region between 1.8\,kpc and 6.8\,kpc.
The interior of the main disc slightly overlaps with the outermost limits of the inner disc.
Both the stellar and dust distribution are relatively flat in the main disc, with a scale length of 4.3\,kpc for the young stellar disc and 6\,kpc for the dust disc.
The main disc contributes most significantly to the global SFR, with a SFR$^{\text{m}}=2.73^{+0.28}_{-0.26}\,$\mpyr.
The main disc has a surface density SFR of $\Sigma_{\text{SFR}}^{\text{m}}=2.00^{+0.18}_{-0.17}\times10^{-2}\,$\mpyrkpc.
The main disc has a dust mass of M$_{\text{d}}^{\text{m}}=3.10^{+0.26}_{-0.18}\times10^{7}\,\text{M}_{\odot}$, with a dust temperature of 21.8\,K.

\subsection{The contribution of young and old stellar populations in heating the dust}
The fractional contribution of the different stellar populations in heating the dust is an important quantity, since it allows a quantitative understanding of the physical mechanisms related to the interaction between stellar light and interstellar dust. In addition, this is key to the understanding of key correlations, like for example the tight and universal correlation between far-infrared and radio continuum emission of spiral galaxies \citep{deJong1985, Helou1985, Wunderlich1987}. The standard interpretation of the so-called FIR-radio correlation is in terms of young and massive star formation activity \citep{Condon1992}. This picture assumes that the dust emission is mainly powered by young stars which are also responsible for the radio emission: the ionising radiation from the  young stars powers the thermal radio emission, and the remnants of the supernova explosions which occur at the end of their lives accelerate the cosmic ray electrons.
Nonetheless both the effect of dust opacity and the contribution of the old stellar populations play a role in shaping the slope of the correlation  \citep{Pierini2003}, and as such, knowledge of the fraction of dust heating powered by the young and old stellar populations is a pre-requisite for the analysis of the FIR/radio correlation.
Furthermore, knowledge of the different stellar populations in heating the dust is equivalent to understanding the dust attenuation of the different stellar populations, which, in turn, allows population synthesis models to be used for deriving star formation histories in galaxies.

 For the global emission of M51 we found that the fractional contribution of the young stellar population in powering the dust emission, $F^{\rm dust}_{\rm young}$, is $69\%$, the rest being attributed to the old stellar populations in the disc and bulge. Previous modelling of M51 with radiative transfer methods derived $F^{\rm dust}_{\rm young}=0.63$ \citep{DeLooze2014} and $F^{\rm dust}_{\rm young}=0.72$ \citep{Nersesian2020b}. The latter value is more in range with our determination.
 
 Our value of $F^{\rm dust}_{\rm young}=0.69$ is similar to that derived with the same models for other nearby spirals: $F^{\rm dust}_{\rm young}=0.71$ for the Milky Way \citep{Natale2021} and  NGC~628 \citep{Rushton2022}, but slightly lower than $F^{\rm dust}_{\rm young}=0.8$ for M33 \citep{Thirlwall2020}.

\begin{figure}
    \centering
    \includegraphics[width=\columnwidth]{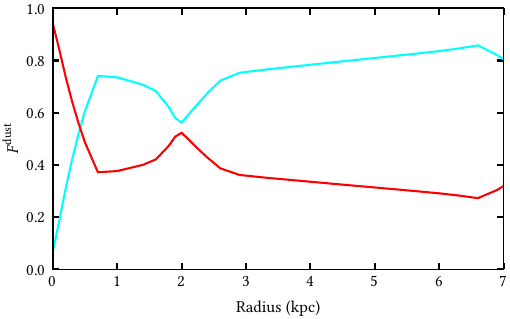}
    \caption{Radial profile of the fraction of dust heating $F^{\rm dust}$ powered by the young (blue) and old (red) stellar populations.}
    \label{fig:frac_eabs}
\end{figure}

Following the formalism from \cite{Natale2015} and \cite{Popescu2000}, we use the radiative transfer calculations to not only derive $F^{\rm dust}_{\rm young}$ and $F^{\rm dust}_{\rm old}$   for the global emission, but also to understand the spatial variation of this quantity. In Fig.~\ref{fig:frac_eabs} we plot $F^{\rm dust}_{\rm young}$ and $F^{\rm dust}_{\rm old}$ as a function of radial distance. Within the inner 0.5 kpc dust emission is dominated by the old stellar populations, mainly from the bulge. Beyond 0.5\,kpc it is the young stellar population that contributes a significant fraction of the dust heating. At around 0.7-0.8\,kpc (the inner radius of the inner disc) there is a peak in $F^{\rm dust}_{\rm young}$, reaching above $70\%$. Within the extent of the main disc, between 2.7 and 7 kpc, $F^{\rm dust}_{\rm young}$ is slowly increasing with radial distance, from $60$ to $85\%$. At around 2\,kpc, corresponding to the transition region between the inner and the main disc, 
$F^{\rm dust}_{\rm young}\approx F^{\rm dust}_{\rm old}$.

The overall trends are similar to those found for the Milky Way \citep{Natale2021} and for NGC~628 \citep{Rushton2022}, in the sense that dust heating in the inner regions has a significant contribution from old stellar populations, while the rest of the dust in the disc is mainly heated by young stellar populations, with $F^{\rm dust}_{\rm young}$ rather constant or slowly increasing towards the outer disc. M51 has larger radial variations in $F^{\rm dust}$ than the Milky Way or NGC~628.

\subsection{The radial variation of \texorpdfstring{$\Sigma_{\text{SFR}}$, $\Sigma_{\text{M}_{\star}}$} and sSFR}

\begin{figure}
    \centering
    \includegraphics[scale=0.6]{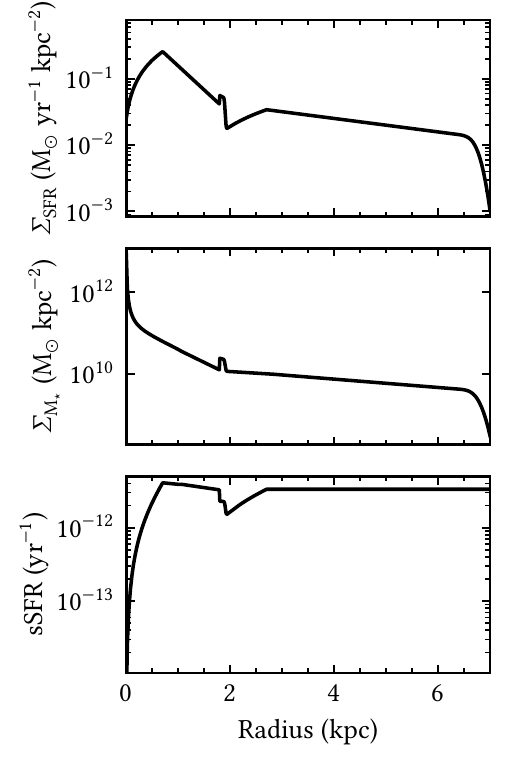}
    \caption{Radial profiles of the SFR surface density. $\Sigma_{\text{SFR}}$ (top), stellar mass surface density, $\Sigma_{\text{M}_{\star}}$ (middle), and specific star formation rate, sSFR (bottom).}
    \label{fig:rad_props}
\end{figure}

The global SFR of M51 was found to be $4.11\,\text{M}_{\odot}\,{\rm yr}^{-1}$. The rate of star formation is not constant throughout the extent of M51, but has a strong radial dependence. To illustrate this we plot in the top panel of Fig.~\ref{fig:rad_props} the surface density of SFR, $\Sigma_{\text{SFR}}$, versus radial distance. One can see that there is a peak in $\Sigma_{\text{SFR}}$ at the inner radius of the inner disc, followed by a steep decrease out to around 2 kpc. Beyond 3 kpc, within the extent of the main disc, the $\Sigma_{\text{SFR}}$ follows a shallow decrease. Overall, $\Sigma_{\text{SFR}}$ shows a marked difference between the inner and the main disc. By contrast, the surface density of stellar mass, 
$\Sigma_{\text{M}_{\star}}$, has a smoother radial variation, (middle panel, Fig.~\ref{fig:rad_props}), showing a shallow decline with increasing radial distance. The stellar mass was calculated using the 3.6 and 4.5\,${\mu}$m flux densities and the calibration from \cite{Eskew2012}.

The specific SFR, sSFR is rather constant throughout the disc (bottom panel, Fig.~\ref{fig:rad_props}), except for the very central region. These trends are similar to the trends found for M33, NGC~628 and the Milky Way.

\subsection{The attenuation curve of M51}
The attenuation curve of a galaxy depends not only on the optical constants of the dust grains and their grain size distribution (incorporated in the dust extinction), but also on the relative distribution of the stars and dust.
Neither the extinction nor the geometry is usually known, making predictions for attenuation rather difficult. It is however the strength of radiative-transfer modelling to accurately derive geometries of stars and dust, within the framework of a given dust model.

\begin{figure}
    \centering
    \includegraphics[width=\columnwidth]{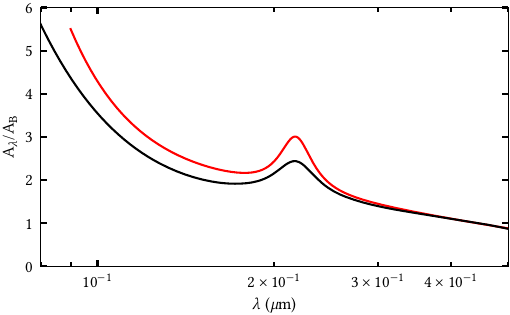}
    \caption{Comparison between the attenuation curve of M51 as seen at an inclination of 20.3$^{\circ}$ (solid red line) with the average extinction curve of the Milky Way (solid black line) \citep{Fitzpatrick1999}.
    Both curves are normalised to the corresponding values in the B-band.}
    \label{fig:atten}
\end{figure}

In Fig.~\ref{fig:atten} we plot the predicted attenuation curve of M51, as derived from our radiative transfer model.
For comparison we also plot the extinction curve of the MW from \cite{Fitzpatrick1999}.
We fit the attenuation curve of our model with a third-order polynomial plus a Drude profile, which is the functional form  presented in \cite{Salim2018}. 
Using this form, we found that the best fit is given by:

\begin{align}
    k_{\lambda} &= -4.95+2.00\lambda^{-1}-0.29\lambda^{-2}+0.02\lambda^{-3}+D_{\lambda}+4.00 \\
    D_{\lambda} &= \frac{3.00\,\lambda^{2}(0.035\,\mu \text{m})^{2}}{[\lambda^{2}-(0.2175\,\mu \text{m})^{2}]^{2}+\lambda^{2}(0.035\,\mu \text{m})^{2}}
\end{align}

The attenuation curve of M51 is slightly steeper than the MW extinction curve, becoming increasingly disparate at shorter wavelengths.
M51 also exhibits a much greater 2200\,\AA\, bump compared to the 
MW extinction.
We find that the difference in width of the bumps to be negligible.

\section{Discussion}
\label{section:discussion}
\begin{figure*}  
    \centering\includegraphics[scale=4.3]{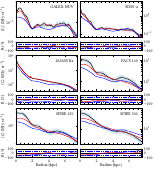}
    \caption{The azimuthally averaged surface-brightness observed profiles (black) plotted against our model (red) and the model of \protect\cite{Nersesian2020b} (blue), for a selection of wavebands.
    The lower panels indicate the residuals for our model (red) and the model of \protect\cite{Nersesian2020b} (blue), with the $\pm 50$ per cent shown with the dashed blue line.}
    \label{fig:nersesian_comparison}
\end{figure*}

\subsection{Comparison with the non axi-symmetric model of Nersesian et al. 2020b} 

As mentioned before, M51 was also modelled with non axi-symmetric RT codes by \cite{DeLooze2014}  using the SKIRT code \citep{Baes2011,Camps2020}, and more recently by \cite{Nersesian2020b}, using a more up to date version of the same formalism and code. Here we compare our model to that of \cite{Nersesian2020b}. Publicly available model images of \cite{Nersesian2020b} were extracted from DustPedia\footnote{https://sciences.ugent.be/skirtextdat/DustPediaData/M51/mock\_images.zip} and processed in the same way as our model images, to produce azimuthally-averaged radial profiles. We also checked that the observed fluxes used in \cite{Nersesian2020b} were consistent with the values used in our work, and, when small differences existed (at percent level), we adjusted for this. In Fig.~\ref{fig:nersesian_comparison} we show examples of this comparison for a few selected wavelengths. Overall our model performs better. This is to be expected, as the model of \cite{Nersesian2020b} does not fit the geometry of the system, but only the spatially integrated SED.

\subsection{An outer disc}
{All the results presented so far were confined to the main extent of the galaxy, out to the radial distance $R_{\rm l}=7$\,kpc. Note however that $R_{\rm l}$ is not the actual truncation radius of M51. There is very faint and asymmetric emission beyond $R_{\rm l}$, also containing the bridge between M51 and its companion M51b. This faint emission does not obey axi-symmetry, therefore it is not a prime target for an axi-symmetric model. Nonetheless, the emission is so faint that it is difficult to trace and model, and one effective way to reveal it is through azimuthally averaging. By averaging, a strong signal becomes apparent,  in a form of an exponential decay. We model this outer emission with another morphological component that we call \lq\lq outer disc\rq\rq\,. While we acknowledge that an axi-symmetric model is a poor approximation for this emission, we nevertheless attempt to at least give some characterisation of the global properties of this emission. 

The model for the outer disc morphology contains the same geometrical components as for the main disc: a stellar disc, a thin stellar disc, a dust disc and a thin dust disc. The analytic functions we fit are the same, and the whole optimisation follows the same procedure as for the main body of the galaxy.

\begin{figure}
    \centering
    \includegraphics[width=\columnwidth]{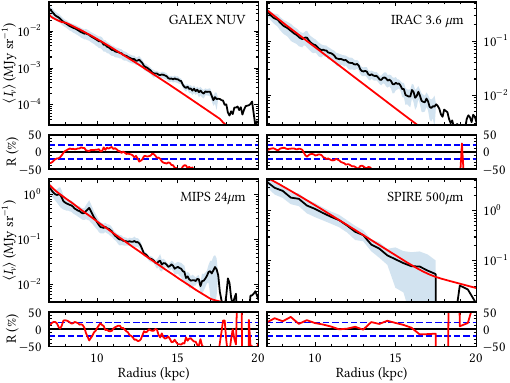}
    \caption{Fits to the surface brightness profiles of the outer disc (7-20\,kpc) at selected wavelengths, chosen to show examples of dust attenuated stellar light in the NUV and 3.6\,${\mu}$m and dust emission in the 24 and 500\,${\mu}$m. The black line is for the observed profiles and the red line is for the corresponding model. The bottom panels show residuals between observations and model.}
    \label{fig:sb_outer}
\end{figure}

Examples of fits to the outer disc are given in Fig.~\ref{fig:sb_outer}. We find the outer disc to extend from 6.7\,kpc to 20\,kpc, with a scale length of 3.6\,kpc and 1.45\,kpc for the dust disc and young stellar disc, respectively. The main geometrical parameters of the outer disc are listed in Table~\ref{tab:geometry_outer}.

The outer disc is the morphological component with the smallest SFR, with a SFR$^{\text{o}}=0.57\pm0.06\,$\mpyr.
The surface density of SFR is the least dense of the components, with $\Sigma_{\text{SFR}}^{\text{o}}=5.11\pm0.56\times10^{-4}\,$\mpyrkpc.

\begin{table}[h]
\centering
\begin{tabular}{l|l}
    \hline
    $h_{\text{s}}^{\text{o-disc}}$\text{(g, r, i J, K, I1, I2, I3)} & (2.0, 2.0, 2.0, 1.72,\\
     & 1.77, 1.95, 1.98, 2.00)$\pm$13\% \\
    \\
    $h_{\text{s}}^{\text{o-tdisc}}$ & 1.45$\pm$0.2 \\
    \\
    $h_{\text{d}}^{\text{o-disc}}$ & 3.6$^{+1.4}_{-1.0}$ \\
    \hline
\end{tabular}
\caption{Main geometric parameters of the outer disc}
\label{tab:geometry_outer}
\end{table}

\subsection{A model with a spatial variation in dust properties}
Inspection of the fitted profiles from Figs.~\ref{fig:wd01_uv}-\ref{fig:wd01_dust} shows that the axi-symmetric model provides a reasonable fit to the data, within $R_l$.

However, when deriving the spatially integrated SED (see Fig.~\ref{fig:8um}, top left-hand panel), we noticed that our predicted intrinsic (de-attenuated) stellar SED shows a spike in the NUV, exactly at the position of the well-known \lq\lq2200\,\AA\,bump\rq\rq\  of the extinction curve.
It can be seen that the inner disc contributes most significantly to this spike, while the main disc has negligible contribution, if any. The intrinsic stellar SED should be smooth, with no feature at this wavelength.
Because our model is in good agreement with the data at all wavelengths, as seen in Figs.~\ref{fig:wd01_uv}-\ref{fig:wd01_dust}, the spike in the intrinsic SED  is indicative that the inner disc is being too heavily attenuated in the NUV. This could be due to a spatial variation in the dust properties, in the sense 
 that the dust properties may be different in the inner disc than assumed in our model.
As mentioned before, our model for M51 uses the Milky Way-type dust from \cite{WeingartnerDraine2001}, described on line 7 in Table 1 from \cite{WeingartnerDraine2001}.
The extinction curve of the Milky Way dust model along with the contributions from the different grain types is shown in the left panel of Fig.~\ref{fig:ext_both}, together with its prominent \lq\lq2200\,\AA\,bump\rq\rq.
This feature is primarily due to the PAH molecules, which the model considers to be the carbonaceous grains with sizes $a\leq 0.01\,\mu$m.
The over-attenuation of the NUV photons in the inner disc could suggest that the extinction curve corresponding to the dust model that we use has a too strong bump, meaning an overabundance of PAH in the inner disc, than is present in M51. We note that it is extremely hard to constrain the shape of the UV extinction curve of galaxies, especially in the inner region where the attenuation is high and the geometry is complex, with just the GALEX FUV and NUV bands. Several authors have shown that these bands alone are insufficient to disentangle the shape of the continuum and the strength of the 2200A bump \citep{Hoversten2011,Hutton2015,Hagen2017,Decleir2019}.

Nonetheless, within the framework of a model with varying dust properties, our fits at $8\,\mu$m (see Fig.~\ref{fig:8um}, bottom right-hand panel) provide additional evidence that we may need a different type of dust in the inner disc.
The model profile is clearly over-predicted in the inner region.
Since the $8\,\mu$m is a strong tracer of PAH abundance, in addition to tracing the radiation fields that are self-consistently calculated in our RT model, the over-prediction could again be indicative of an overabundance of PAH in our model of the inner disc.

As an alternative, we sought a dust model with a reduced PAH abundance and chose one of the LMC dust models of \cite{WeingartnerDraine2001}, namely the LMC average (line 1 from Table 3 of \citealp{WeingartnerDraine2001}).
This LMC-type dust features the same optical constants but different grain size distributions, particularly less PAH grains.
Thus, the extinction curve exhibits a less pronounced 2200\,\AA\,bump, as seen in the right panel of Fig.~\ref{fig:ext_both}.
As the attenuation in the main and outer discs produced smooth intrinsic SEDs, we chose to modify the dust properties only in the inner disc (using the LMC-type dust). We thus fitted a new model of M51 with variable dust grain properties. We refer to this alternative as the \lq\lq hybrid model\rq\rq.
The new fits produce the intended effect, reducing the NUV peak in the de-attenuated stellar SED of the inner disc, as seen in Fig.~\ref{fig:8um}.
The new fits to the $8\,\mu$m profiles (see Fig.~\ref{fig:8um}, bottom right-hand panel) also provide the right level of emission at this wavelength.
We conclude that using the LMC-type dust for the inner disc could provide an alternative to the standard model. 

 The finding that PAHs could be less abundant in the inner disc than in the rest of the galaxy raises the question of what physical processes could cause this variation. Theoretical and observational studies suggest that strong UV radiation fields could destroy PAH molecules (e.g.,~\citealp{Boulanger1988,Helou1991,Contursi2000,Siebenmorgen2010}). Taking into account that the inner region of M51 has the highest surface density of SFR, and therefore intense UV radiation fields, it is plausible that PAHs are more readily destroyed in this region, thus explaining the reduced abundance inferred by the hybrid model. Several studies have suggested that the fraction of PAHs becomes low in HII regions (e.g.,~\citealp{Pety2005,Lebouteiller2007,Thilker2007,Chastenet2023}) and is a strong function of environment, with higher fractions in less harsh environments due to star formation \citep{Chastenet2023}. It is beyond the scope of our paper to provide a quantitative model for PAH destruction in different ISM environments, but we note the possibility of PAH abundance variation within galaxies.

\begin{figure*}
    \centering
    \includegraphics[scale=0.6]{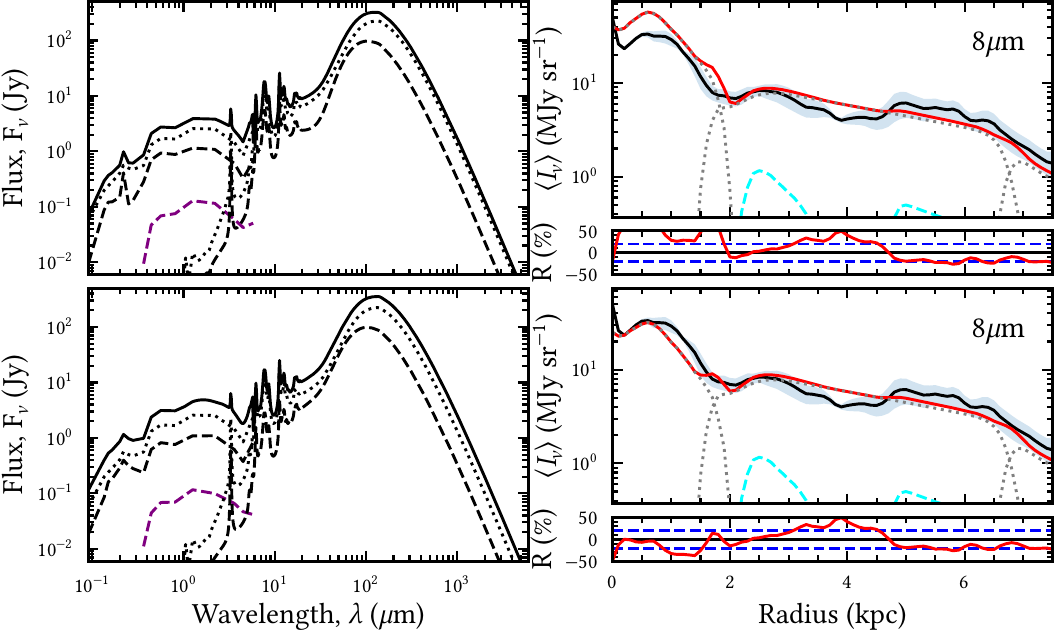}
    \caption{Top panels:  the standard model (constant dust properties) for the intrinsic global SED (top left) and the surface brightness profile at 8\,$\mu$m (top right). Bottoms panels: same but for the hybrid model (with spatially varying dust properties).
    The left hand plots show the intrinsic global SED (solid black line) plotted together with the contribution from the individual morphological  components: the inner (dashed black line) and main (dotted black line) discs as well as the bulge (dashed purple line).
    The right hand plots show the $8\,\mu$m profile of the observation (solid black line) compared to the  model (solid red line) along with the contributions from the dust discs (dotted grey lines), and the HII components (dashed cyan lines). Residuals between observed and model profiles are also shown on the bottom of each panel.}
    \label{fig:8um}
\end{figure*}

The fits to the multiwavelength data using the hybrid model do not provide any big differences from the standard model, with the exception of the 8\,$\mu$m profile, which is largely improved (see examples in the Appendix, in Figs.~\ref{fig:combined_uv}-\ref{fig:combined_dust}).  
The energy balance between direct and re-radiated stellar light seems to be the same as in the standard model, with no major improvements in the fit. The values of the ${\rm chi}^2$ are also similar. 
The individual values at selected wavelengths are listed in Table~\ref{tab:chi_sqr_combined}.
The values of the derived intrinsic parameters are different, but the quality of the fit is the same. This is perhaps not surprising, since the optical properties of the grains are not changed. There are some other dust models available in the literature, in particular "The Heterogeneous dust Evolution Model for Interstellar Solids" (THEMIS) \citep{Jones2013,Koehler2014,Ysard2015}, which consider different carbonaceous grains, with aliphatic rather than aromatic molecules. The THEMIS model has enhanced submm efficiencies for grains, by a factor of about 2.5 with respect to the Draine model. The THEMIS model would produce a different energy balance and therefore different fits. 

To conclude, the hybrid model with a reduced PAH abundance in the inner disc, together with a modified grain size distribution in the central region, provides a consistent solution to the panchromatic data of M51, and is based on observational evidence coming from both dust extinction and dust emission. Nonetheless, we cannot prove that these observational constraints are uniquely met by our hybrid model.

\section{Summary}
\label{sec:summary}
Using the generic radiative transfer code and formalism of \citetalias{Popescu2011} we derived an axisymmetric model for M51, fitted to multiwavelength imaging data ranging from FUV to submm. We find that, despite the interaction with the companion galaxy M51b, M51 preserves a regular spiral structure within a radial distance of 
$R_{\rm l}$=7\,kpc, making it thus suitable for the assumption of axi-symmetry within $R_{\rm l}$. The model fits to the azimuthally-averaged surface brightness profiles were found to be in reasonable agreement with the observations. 

We find three distinct morphological components out to $R_{\rm l}$: a bulge, an inner disc and a main disc. The bulge has an effective radius $R_{\rm eff}=350$\,pc and a S\'ersic index $n=4$. The inner disc is prominent in the UV, with a scale-length of the thin stellar disc, $h_{\rm s}^{\rm i-disc}$, of 600 pc. The distribution of dust in the inner disc is very flat, with a scalelength of the dust disc, $h_{\rm d}^{\rm disc}$, of 5 kpc. The main disc resides in the region between $\sim$ 1.8\,kpc and $\sim$6.8\,kpc. Both the stellar and the dust distributions of this morphological component are relatively flat, with a scale-length for the thin stellar disc, $h_{\rm s}^{\rm t-disc}$ of 4.3\,kpc and $h_{\rm d}^{\rm disc}=6$\, kpc for the dust disc.

The main intrinsic properties of M51, as derived from the model are as follows:

\begin{itemize}
    \item The global star-formation rate  is $\text{SFR}=4.11^{+0.40}_{-0.38}\,$\mpyr. The inner disc has $\text{SFR}=1.38^{+0.15}_{-0.14}\,$\mpyr and the main disc has $\text{SFR}=2.73^{+0.28}_{-0.26}\,$\mpyr.

    \item The global surface density in SFR is $\Sigma_{\text{SFR}}=2.83^{+0.28}_{-0.27}\times10^{-2}\,$\mpyrkpc.
    
    \item The specific star formation rate, sSFR, is rather constant over much of the extent of M51, except for the very centre.
 
    \item The young stars account for 69\% of the dust heating. In the inner regions dust heating has a significant contribution from old stellar populations, while the rest of
the dust in the disc is mainly heated by young stellar populations, with their contribution rather constant or slowly increasing towards
the outer disc.
\end{itemize}

We discuss the properties of the faint and non axi-symmetric emission extending beyond $R_{\rm l}$, also containing the bridge between M51 and its companion M51b.  We model this outer emission with another
morphological component that we call \lq\lq outer disc\rq\rq\,. 
The outer disc has $\text{SFR}=0.57\pm0.06\,$\mpyr and $\Sigma_{\text{SFR}}=5.11\pm0.56\times10^{-4}\,$\mpyrkpc.

We also show a model with varying dust properties, with the inner disc having a reduced PAH abundance with respect to the main disc.  This hybrid model was chosen to solve the over-prediction of the standard model at 8\,$\mu$m in the inner disc, and to alleviate a residual bump in the derived intrinsic SED in the NUV band, at the position of the 2200\,\AA bump. 

\section*{Acknowledgements}
We would like to thank an anonymous referee for very useful and constructive comments, that helped improve the manuscript.
This work is based in part on observations made with the NASA Galaxy Evolution Explorer. GALEX is operated for NASA by the California Institute of Technology under NASA contract NAS5-98034.  
This research has made use of the NASA/IPAC Infrared Science Archive, which is operated by the Jet Propulsion Laboratory, California Institute of Technology, under contract with the National Aeronautics and Space Administration.

This work is also based on the Sloan Digital Sky Survey (SDSS) data.
Funding for the SDSS IV has been provided by the Alfred P. Sloan Foundation, the U.S. Department of Energy Office of Science, and the Participating Institutions. SDSS acknowledges support and resources from the Center for High-Performance Computing at the University of Utah. The SDSS web site is www.sdss4.org.
SDSS is managed by the Astrophysical Research Consortium for the Participating Institutions of the SDSS Collaboration including the Brazilian Participation Group, the Carnegie Institution for Science, Carnegie Mellon University, Center for Astrophysics | Harvard \& Smithsonian (CfA), the Chilean Participation Group, the French Participation Group, Instituto de Astrofísica de Canarias, The Johns Hopkins University, Kavli Institute for the Physics and Mathematics of the Universe (IPMU) / University of Tokyo, the Korean Participation Group, Lawrence Berkeley National Laboratory, Leibniz Institut für Astrophysik Potsdam (AIP), Max-Planck-Institut für Astronomie (MPIA Heidelberg), Max-Planck-Institut für Astrophysik (MPA Garching), Max-Planck-Institut für Extraterrestrische Physik (MPE), National Astronomical Observatories of China, New Mexico State University, New York University, University of Notre Dame, Observatório Nacional / MCTI, The Ohio State University, Pennsylvania State University, Shanghai Astronomical Observatory, United Kingdom Participation Group, Universidad Nacional Autónoma de México, University of Arizona, University of Colorado Boulder, University of Oxford, University of Portsmouth, University of Utah, University of Virginia, University of Washington, University of Wisconsin, Vanderbilt University, and Yale University.

This work has also made use of data products from the Two Micron All Sky Survey, which is a joint project of the University of Massachusetts and the Infrared Processing and Analysis Center/California Institute of Technology, funded by the National Aeronautics and Space Administration and the National Science Foundation. 
This work is based in part on observations made with the {\it Spitzer} Space Telescope, which is operated by the Jet Propulsion Laboratory, California Institute of Technology under a contract with NASA.
We also utilise observations performed with the ESA {\it Herschel} Space Observatory \citep{Pilbratt2010}, in particular to do photometry using the PACS  \citep{Poglitsch2010} and SPIRE \citep{Griffin2010} instruments.
\newpage
%%%%%%%%%%%%%%%%%%%%%%%%%%%%%%%%%%%%%%%%%%%%%%%%%%
\section*{Data Availability}

%The data underlying this article will be shared on reasonable request to the corresponding author.
The data underlying this article are made available at the CDS database via http://cdsweb.u-strasbg.fr/cgi-bin/qcat?J/MNRAS/



\bibliographystyle{mnras}
\bibliography{./References.bib} % if your bibtex file is called 

%%%%%%%%%%%%%%%%%%%%%%%%%%%%%%%%%%%%%%%%%%%%%%%%%%

%%%%%%%%%%%%%%%%% APPENDICES %%%%%%%%%%%%%%%%%%%%%
\newpage
\appendix
%\onecolumn

\section{Amplitude and geometrical parameters}
\label{Appendix:results-tables}
In our model we express the amplitude parameters of the stellar discs with  spectral luminosity densities $L_{\nu}$, and the dust discs with central face-on optical depths in the B-band, $\tau_\text{B}$.
Table~\ref{tab:int_lum_wd01} lists the spectral luminosity densities for each stellar disc and the bulge.
Table~\ref{tab:optical depth} lists the B-band face-on optical depth for each of the morphological components.
Table~\ref{tab:geometry regions} lists other parameters, like the $R_{\text{in}}$, $R_{\text{tin}}$ and $R_{\text{t}}$, which are fixed from data. The table also lists the $\chi$ values for the main disc. The latter parameter serves to more cohesively join the inner to the main disc.

\begin{table}[htbp]
\small
\setlength{\tabcolsep}{3pt}
\centering
\caption{The intrinsic spectral luminosity densities of the stellar and thin stellar disc, for each morphological component.}
\label{tab:int_lum_wd01}
\begin{tabular}[t]{c|c|c|c|c|c}
\hline
$\lambda$ & $L_{\nu}^{\text{bulge}}$ & $L_{\nu}^{\text{disc, i}}$ & $L_{\nu}^{\text{disc ,m}}$ & $L_{\nu}^{\text{tdisc, i}}$ & $L_{\nu}^{\text{tdisc, m}}$ \\
(\AA) & (W Hz$^{-1}$) & (W Hz$^{-1}$) & (W Hz$^{-1}$) & (W Hz$^{-1}$)& (W Hz$^{-1}$) \\
\hline
1542 & - & - & - & 1.07$\times10^{21}$ & 2.11$\times10^{21}$ \\
2274 & - & - & - & 3.62$\times10^{21}$ & 4.97$\times10^{21}$ \\
3562 & 8.31$\times10^{19}$ & 7.42$\times10^{20}$ & 1.48$\times10^{21}$  & 1.75$\times10^{21}$ & 3.42$\times10^{21}$ \\
4719 & 4.16$\times10^{20}$ & 4.36$\times10^{21}$ & 8.20$\times10^{21}$  & 1.87$\times10^{21}$ & 3.39$\times10^{21}$ \\
6185 & 6.05$\times10^{20}$ & 6.26$\times10^{21}$ & 1.21$\times10^{22}$  & 1.96$\times10^{21}$ & 2.95$\times10^{21}$ \\
7500 & 6.64$\times10^{20}$ & 6.08$\times10^{21}$ & 1.24$\times10^{22}$  & 1.66$\times10^{21}$ & 2.26$\times10^{21}$ \\
12000 & 1.13$\times10^{21}$ & 9.26$\times10^{21}$ & 2.11$\times10^{22}$  & 8.17$\times10^{20}$ & 1.48$\times10^{21}$ \\
22000 & 1.02$\times10^{21}$ & 8.94$\times10^{21}$ & 2.12$\times10^{22}$  & 6.40$\times10^{20}$ & 1.16$\times10^{21}$ \\
35070 & 5.25$\times10^{20}$ & 5.25$\times10^{21}$ & 1.23$\times10^{22}$  & 1.25$\times10^{20}$ & 5.63$\times10^{20}$ \\
44370 & 3.73$\times10^{20}$ & 2.75$\times10^{21}$ & 6.86$\times10^{21}$  & 1.25$\times10^{20}$ & 5.63$\times10^{20}$ \\
57390 & 4.47$\times10^{20}$ & 5.74$\times10^{21}$ & 2.05$\times10^{22}$  & 1.50$\times10^{20}$ & 5.63$\times10^{20}$ \\
\hline
\end{tabular}
\end{table}

\begin{table}[htb]
    \centering
    \caption{The face-on optical depth in the B-band at the inner radius of their respective morphological component.}
    \begin{tabular}{c|c}
         & $\tau_{\text{B}}^{\text{f}}\left(R_{\text{in}}\right)$ \\
        \hline
        Inner & 5.33$\pm0.33$ \\
        Main & 2.62$\pm0.08$ \\
        \hline
    \end{tabular}
    \label{tab:optical depth}
\end{table}

\begin{table}[H]
\centering
\begin{tabular}{l|l}
    \hline
    Fitting parameters \\
    \hline
    $\chi_{\text{s}}^{\text{m-tdisc}}$ & -0.7$\pm0.5$ \\
    $\chi_{\text{s}}^{\text{m-disc}}$ & 1.5$\pm0.25$ \\
    $\chi_{\text{d}}^{\text{m-disc}}$ & 1$\pm0.25$  \\
    \hline
    Parameters fixed by data \\
    \hline
    $R_{\text{in,s}}^{\text{(i-tdisc, m-tdisc)}}$ & (0.7, 2.7) \\
    $R_{\text{in,s}}^{\text{(i-disc, m-disc)}}$ & (0, 2.7) \\
    $R_{\text{in,d}}^{\text{(i-disc, m-disc)}}$ & (0.8, 2.7) \\
    \\
    $R_{\text{tin,s}}^{\text{(i-tdisc, m-tdisc)}}$ & (0, 1.8) \\
    $R_{\text{tin,s}}^{\text{(i-disc, m-disc)}}$ & (0, 1.8) \\
    $R_{\text{tin,d}}^{\text{(i-disc, m-disc)}}$ & (0, 1.8) \\
    \\
    $R_{\text{t,s}}^{\text{(i-tdisc, m-tdisc)}}$ & (1.9, 6.8) \\
    $R_{\text{t,s}}^{\text{(i-disc, m-disc)}}$ & (1.9, 6.8) \\
    $R_{\text{t,d}}^{\text{(i-disc, m-disc)}}$ & (1.9, 6.8) \\
    \hline
    Parameters fixed from model\\
    \hline
    $z_{\text{s}}^{\text{(i-tdisc, m-tdisc)}}$ & (0.09, 0.09) \\
    $z_{\text{s}}^{\text{(i-disc, m-disc)}}$ & (0.19, 0.19) \\
    $z_{\text{d}}^{\text{(i-disc, m-disc)}}$ & (0.16, 0.16) \\
    
    \hline
\end{tabular}
\caption{Other parameters of the model. All length parameters are given in kpc.}
\label{tab:geometry regions}
\end{table}

\newpage
\onecolumn
\section{Star formation and dust masses}
In this appendix we present tables with the calculated values for SFR (Table.~\ref{tab:sfr}), $\Sigma_{\text{SFR}}$ (Table.~\ref{tab:surface_sfr}), and  dust mass (Table.~\ref{tab:dust_mass}).
All these values are given for both the global emission and for each morphological component.

\begin{table}[H]
    \centering
    \caption{Star formation rates for the global emission and for the morphological components of M51. Values are in (M$_{\odot}$\,yr$^{-1}$).}
    \label{tab:sfr}
    \begin{tabular}{c|c|c}
         & SFR (M$_{\odot}$yr$^{-1}$)  \\
        \hline
        Global & 4.11$^{+0.41}_{-0.39}$\\
        Inner & 1.38$^{+0.15}_{-0.14}$\\
        Main & 2.73$^{+0.28}_{-0.26}$\\
        \hline
    \end{tabular}
\end{table}

\begin{table}[H]
    \centering
    \begin{tabular}{c|c|c}
         & $\Sigma_{\text{SFR}}$ (M$_{\odot}$yr$^{-1}$kpc$^{-2}$) \\
        \hline
        Global & 2.83$^{+0.28}_{-0.27}\times10^{-2}$\\
        Inner & 1.21$^{+0.12}_{-0.11}\times10^{-1}$\\
        Main & 2.00$^{+0.18}_{-0.17}\times10^{-2}$\\
        \hline
    \end{tabular}
    \caption{Surface density of star formation for the global emission and for the morphological components of M51. Values are in (M$_{\odot}$\,yr$^{-1}$\,kpc$^{2}$).}
    \label{tab:surface_sfr}
\end{table}

\begin{table}[H]
    \centering
    \begin{tabular}{c|c|c}
         & dust mass M$_{\text{d}}$ (M$_{\odot}$) \\
        \hline
        Global & 3.74$^{+0.34}_{-0.30}\times10^{7}$\\
        Inner & 6.39$^{+0.37}_{-0.21}\times10^{6}$\\
        Main & 3.10$^{+0.26}_{-0.18}\times10^{7}$\\
        \hline
    \end{tabular}
    \caption{Dust masses for the global emission and for the morphological components of M51. Values are in (M$_{\odot}$).}
    \label{tab:dust_mass}
\end{table}

\begin{table}[H]
    \centering
    \begin{tabular}{c|c}
    \hline
        Band & $\text{chi}^{2}_{r}$ \\
        \hline
        NUV & 0.97 \\
        I2 & 2.60 \\
        PACS70 & 2.01 \\
        SPIRE500 & 0.640\\
        \hline
        Global & 1.89
    \end{tabular}
    \caption{The $\text{chi}^{2}_{r}$ values at selected wavelengths, as well as the total over all wavelengths for the hybrid model.}
    \label{tab:chi_sqr_combined}
\end{table}

\vspace{5cm}
\section{Extinction curves of the Milky Way and LMC dust models}
\label{appendix:dust_models}
In this paper we use the Milky Way and LMC dust models given in \cite{WeingartnerDraine2001} in  Table 1, line 7 and Table 3, line 1 respectively.
Fig.~\ref{fig:ext_both} shows the extinction curve for each grain composition as well as the total extinction curve for the Milky Way and LMC dust models.
These plots demonstrate that the LMC type dust exhibits a reduced 2200\,\AA\, bump compared to the Milky Way type dust.
Figs.~\ref{fig:combined_uv}-\ref{fig:combined_dust} show the fits at various wavelengths using the hybrid model.

\begin{minipage}{\linewidth}
\vspace{0.5cm}
\centering
\includegraphics[width=\columnwidth]{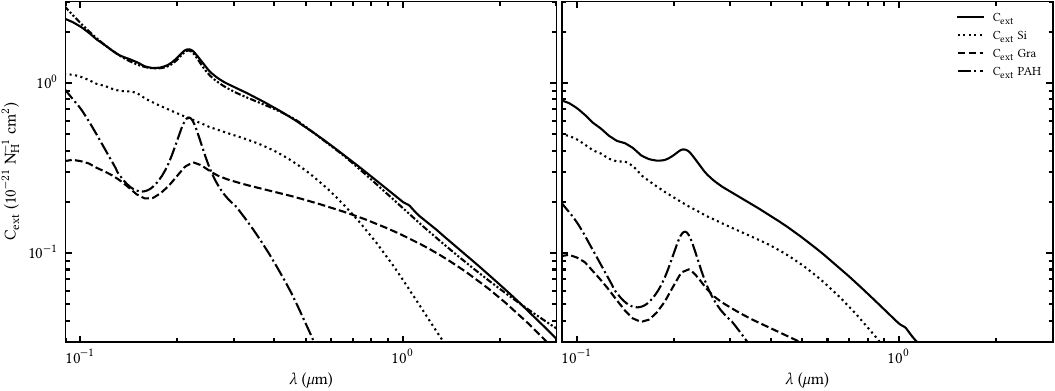}
\captionof{figure}
{Extinction curves for the Milky Way (left) and the LMC (right) dust models, with the grain size distribution and optical constants from \protect\cite{WeingartnerDraine2001}.
We also plot the contributions from the different grain compositions: Si (dotparted), Gra (dashed) and PAH (dot-dashed) as well as the mean extinction curve for the Milky Way (double dot-dashed) \protect\citep{Fitzpatrick1999}.
It can be seen that the LMC dust model features a reduced 2000 bump wh 
It can be seen that the LMC dust model features a reduced 2200\,\AA\, bump when compared to the Milky Way dust.}
\label{fig:ext_both}
\end{minipage}

\begin{figure}
    \centering
    \includegraphics[width=\columnwidth]{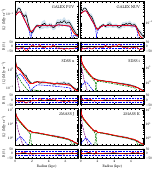}
    \caption{Comparison between the azimuthally averaged surface brightness profiles of the observations (solid black line) at selected UV/optical/NIR wavelengths and of the corresponding model for dust attenuated stellar light (solid red line) using the combined dust models of Milky Way and LMC \citep{WeingartnerDraine2001}. 
    The blue shaded region around the observed profile represents the corresponding errors in the averaged surface brightness, as described in Sect.~\ref{section:data}.
    The contribution from each morphological component is plotted with dashed lines and  colour-coded as follows: bulge in purple, stellar disc in green and thin stellar disc in blue.
    The lower panels show the residuals between the observations and our model, with the dashed blue lines showing the $\pm20\%$ residuals to guide the eye.}
    \label{fig:combined_uv}
\end{figure}
\begin{figure}
    \centering
    \includegraphics[width=\columnwidth]{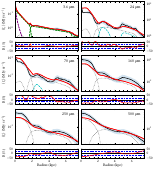}
    \caption{Same as Fig.~\ref{fig:combined_uv} but for selected wavelengths in the NIR/MIR/FIR/submm, where the emission from the dust discs is plotted with the grey dotted lines and the dust emission from the HII regions is shown by the dashed cyan lines.}
    \label{fig:combined_dust}
\end{figure}
%%%%%%%%%%%%%%%%%%%%%%%%%%%%%%%%%%%%%%%%%%%%%%%%%%

% Don't change these lines
\bsp	% typesetting comment
\label{lastpage}
\end{document}